%% file: civhosts.tex
\documentclass[a4paper,fleqn,usenatbib]{mnras}

\usepackage{newtxtext,newtxmath}

\usepackage[T1]{fontenc}
\usepackage{ae,aecompl}

\usepackage{epsf}
\usepackage{graphicx}	
\usepackage{amsmath}	

\input mymacros_nojournals

\newcommand{\fescmax}{f_\mathrm{esc,max}}
\newcommand{\td}{\textit{Technicolor Dawn}}

\newcommand{\tla}{T_{\mathrm{Ly}\alpha}}
\newcommand{\xiga}{\xi_\mathrm{g-abs}}
\newcommand{\xiion}{\xi_\mathrm{ion}}

\title[CIV Hosts]{The Faint Host Galaxies of $\civ$ Absorbers at $z>5$}

\author[K.\ Finlator et al.]{
\parbox[t]{\textwidth}{\vspace{-1cm}
Kristian Finlator$^{1,2,7}$, 
Caitlin Doughty$^{1}$,
Zheng Cai$^{3}$,
Gonzalo D{\'{\i}}az$^{4,5,6}$
}
\\\\$^1$ New Mexico State University, Las Cruces, NM, USA
\\$^2$ Cosmic Dawn Center (DAWN), Niels Bohr Institute, University of Copenhagen / DTU-Space, Technical University of Denmark
\\$^{3}$ Department of Astronomy and Center for Astrophysics, Tsinghua University, Beijing, 100084, China
\\$^{4}$Gemini Observatory, Southern Operations Center, La Serena, Chile
\\$^{5}$Instituto de Ciencias Astron\'omicas, de la Tierra y del Espacio (ICATE), San Juan, Argentina
\\$^{6}$Consejo de Investigaciones Cient\'ificas y T\'ecnicas (CONICET), San Juan, Argentina
\\$^{7}$ finlator@nmsu.edu
\author[The Faint Host Galaxies of $\civ$ Absorbers at $z>5$]{
K.\ Finlator,
Z.\ Cai,
\& G Diaz
}
}

\date{Accepted XXX. Received YYY; in original form ZZZ}

\pubyear{2019}

\begin{document}
\label{firstpage}
\pagerange{\pageref{firstpage}--\pageref{lastpage}}
\maketitle

\begin{abstract}
We explore the expected galaxy environments of $\civ$ absorbers at $z>5$
using the \td~simulations. These simulations reproduce the observed
history of reionization, the $z\sim6$ galaxy stellar mass function, the
\lya forest transmission at $z>5$, and the $\siiv$ column 
density distribution (CDD) at $z\approx5.5$. Nonetheless, the $\civ$ CDD remains 
underproduced. Comparison with observed $\cii/\siii$ equivalent 
width ratios and the $\cii$ line incidence suggests 
that a low carbon yield accounts for some, but not all, of the $\civ$ 
discrepancy. Alternatively, a density-bounded escape scenario could harden the 
metagalactic ionizing background more dramatically even than binary stellar 
evolution, boosting the $\civ$ CDD into near-agreement with observations. 
In this case galaxies ionize more efficiently and fewer are required to 
host a given high-ionization absorber. Absorbers' environments therefore 
constrain ionizing escape. Regardless of the escape scenario, galaxies 
correlate with $\civ$ absorbers out to 300 proper kpc (pkpc). The correlation 
strengthens independently with galaxy luminosity and $\civ$ column density. 
Around strong systems ($\log(N_\civ/\cm^{-2})>14)$), the 
overdensity of galaxies with $\MUV<-18$ or 
$\log(L_{{\rm Ly}\alpha}/\erg\ {\rm s}^{-1}) > 41.9$ declines from 200--300 within 100 pkpc to 
40--60 within 250 pkpc. The previously-suggested association between strong 
$\civ$ absorbers and Ly$\alpha$ emitters at $z>5$ is not expected. It 
may arise if both populations inhabit large-scale voids, but for different 
reasons. Although most neighboring galaxies are too faint for \emph{HST}, 
\emph{JWST} will, with a single pointing, identify $\sim10$ neighboring 
galaxies per strong $\civ$ absorber at $z>5$. Ground-based tests of these 
predictions are possible via deep surveys for \lya emission 
using integral field units. 
\end{abstract}




\section{Introduction} \label{sec:intro}
Over the past twenty years, increasingly sophisticated surveys have uncovered 
thousands of galaxies that were in place and growing vigorously long before
the Epoch of Cosmological Hydrogen Reionization (EOR) ended. The
overall abundance and spatial distribution of young galaxies has been measured 
at luminosities less than 1\% of
$L_*$~\citep{fink15,bouw15,sant16,drak17,live17,atek18,konn18,yue18,bhat19,khus19,vieu19}, yielding
critical constraints on star formation and feedback at early times. The 
emerging consensus that faint galaxies were abundant during the EOR begs
the question as to what feedback processes regulated their growth, and what, 
if anything, they contributed to cosmological hydrogen reionization. 

Theoretical models indicate that the primary mechanism for regulating the 
growth of galaxies in dark matter halos with masses $M_h > 10^9\msun$ 
is galactic outflows~\citep{scha10,somd15}. Outflows, in turn, 
leave signatures in the circumgalactic medium (CGM) that are sensitive probes
of kinetic and radiative feedback. For example, hydrodynamic simulations have 
shown that metal absorbers are severely underproduced if outflows are 
absent~\citep{oppe08}, that the geometric cross section for neutral hydrogen 
absorption is enhanced by outflows~\citep{fauc15}, that the abundance of 
high-ionization absorbers is sensitive to outflow velocities~\citep{keat16}, 
and that models in which more stars form tend to produce more metals and 
therefore stronger metal absorbers~\citep{rahm16}.

While these ensemble studies leverage well the growing catalog of 
high-ionization EOR absorbers that have been identified over the last decade, 
less progress has been made in understanding the relationship between 
individual EOR galaxies and their respective CGM. What sort of absorbers 
are found near galaxies of differing luminosities, and what sort of galaxies 
are expected near absorbers of differing strengths? How do answers to
these questions constrain kinetic and radiative feedback?

Studies of galaxies' environments at $z=$2--3 have shown that bright 
($\sim L_*$) galaxies possess enriched CGM whose metal column density falls 
smoothly with impact parameter~\citep{stei10}. By contrast, environmental 
studies at $z>5$ suggest that strong $\civ$ absorbers are found 
preferentially around faint galaxies rather than bright ones~\citep{diaz14}. 
Does this apparent conflict reflect the difference between 
selecting galaxy-absorber pairs based on galaxy luminosity at low 
redshift versus $\civ$ column density at high redshift, or does it 
indicate that the characteristic host galaxy of strong $\civ$ absorbers 
evolves with time?

One suggested explanation is that faint galaxies dominated the metagalactic
ionizing ultraviolet background (UVB) during the EOR~\citep{diaz14}. 
As a $\civ$ system's
column density increases with both metallicity and UVB amplitude,
faint galaxies could dominate the environments of strong $\civ$ absorbers
either by ejecting more metals than bright galaxies do, or by releasing more ionizing 
light into their environments. If this interpretation is correct, then it
supports an outsized role for faint galaxies in driving reionization and
UVB evolution. Indeed, encouraging qualitative support for the idea that 
absorbers trace LyC emission was recently presented by~\citet{meye19},
who found evidence for local-scale Lyman-$\alpha$ forest opacity fluctuations 
in the vicinity of strong $\civ$ absorbers. 

The possibility of using the environments of high-ionization metal absorbers 
to trace ionizing flux from faint galaxies represents an intriguing complement 
to existing efforts. By far the most popular current approach involves measuring
the galaxy luminosity function (LF), estimating the overall Lyman continuum (LyC) 
emissivity of all galaxies, and comparing it 
to the predicted recombination rate of the intergalactic medium (IGM). This 
method~\citep[for example,][]{mada96} has been used to show that, subject to 
assumptions regarding the extrapolated abundance of faint galaxies, their 
intrinsic LyC emissivity $\xiion$ (that is, the ratio of the ionizing to
non-ionizing luminosity), the fraction $\fesc$ of LyC light 
that escaped into the IGM, and the overall IGM recombination 
rate~\citep{pawl09a,finl12,jees14}, star formation in young galaxies had 
the potential to drive hydrogen reionization to 
completion~\citep{yan04,robe10,robe15,haar12,bouw16,fink19}.

Whether the values of $\xiion$ and $\fesc$ that result from these analyses
are realistic is more difficult to answer observationally~\citep{elli14}. 
Evidence that young 
galaxies had the potential to sculpt their environments comes from the 
strength of their emission lines, which reflect ionizing light from massive 
young stars that has been re-processed in the interstellar medium 
(ISM)~\citep{brom01}. The recent detection of strong line emission 
from galaxies at $z>4$~\citep{star15,bouw16,smit16,rasa16} constrains 
the parameter combination $\xiion(1-\fesc)$ to be larger than expected 
for active, low-metallicity galaxies, but it does not directly trace 
the amount of ionizing flux escaping into the IGM. 

Measurement of galaxies' non-ionizing ultraviolet continua, when modeled
using stellar population synthesis techniques, can be used to constrain
the product $\xiion\fesc$~\citep{dunc15,chish19}. Consistent with other 
studies, these efforts support the possibility that early galaxies packed 
sufficient firepower to complete reionization.  However, results still depend 
on an extrapolation 
from the non-ionizing to the ionizing stellar continuum, which is in
turn sensitive at the $\sim$factor-of-two level to uncertainties in
the underlying stellar populations.

At $z<4$, the relatively-transparent IGM allows direct detection of LyC
flux~\citep{inou08}, yielding a more direct constraint on the product
$\xiion\fesc$~\citep[][and references therein]{rigb19}. In a 
comprehensive analysis of
ground-based spectroscopic measurements of bright galaxies at 
$z\sim3$,~\citet{stei18} report a characteristic escape fraction of 
$\fesc=9\pm1\%$. A central limitation in these results is that the
galaxies at $z\leq4$ for which leaking ionizing flux is directly 
detected may not be representative of the faint systems that dominated 
the UVB at $z\geq5$, particularly if $\fesc<5\%$~\citep{fink19}. 

The idea that bright galaxies, even if somewhat leaky, may not dominate 
the UVB is further underscored by~\citet{kaki18}, who detected a 
statistical association between transparent regions in the \lya forest 
(LAF) and Lyman Break galaxies (LBGs) at $5.3 \lesssim z \lesssim 6.4$.
They find 
that, while the LBGs themselves cannot provide the necessary flux to 
ionize the local LAF, faint galaxies that are presumably clustered 
about them may be able to provided that $\fesc\geq8\%$. This pathbreaking study 
provides independent support for the view that, at $z>5$, the LAF opacity
reflects local-scale UVB fluctuations~\citep[see also][]{davi18c,beck18,kash19}. 
Nonetheless, its result is qualitatively similar to the overall one: an 
unseen population of galaxies with unknown LyC emissivity must be 
invoked in order to explain the observed properties of the high-$z$ LAF.

The escape fraction $\fesc$ from faint galaxies may be faithfully sampled 
via followup spectroscopy of long-duration gamma-ray bursts (GRBs), which 
are associated with core-collapse supernavae~\citep{hjor03}.~\citet{tanv19}, 
applying a 
method developed by~\citet{cpg07}, have shown that GRB spectra inevitably 
show evidence for proximate damped \lya absorbers (DLAs), which 
are optically thick to LyC. They estimate a mean $\fesc$ from the regions 
where GRBs originate of much less than 1\%.  If this number applies 
generally to star-forming regions in the EOR, then not only does it
conflict with studies that directly detect an association between 
galaxies, absorbers, and the LAF~\citep{kaki18,meye19}, it rules out the 
galaxy-driven reionization hypothesis.

In this work, we explore how deep galaxy surveys near strong $\civ$ 
absorbers trace the release of metals and ionizing flux from faint galaxies.
As a by-product, we will show that the next generation of deep followup
surveys using Integral Units as well as the 
\emph{James Webb Space Telescope} (\emph{JWST}) 
will uncover faint galaxies $\sim100\times$ more efficiently than blank-field
surveys when they 
target the environments of strong metal absorbers. While the actual source 
densities in these areas will be biased, they will nonetheless probe the 
faint end of the overall LF indirectly through comparison 
with models that treat the absorber-galaxy relationship realistically.

In Section~\ref{sec:sims}, we review our simulations. In Section~\ref{sec:newResults},
we highlight improvements
with respect to our previous work through comparisons between predictions and
observations of the galaxy stellar mass function, the history of reionization, 
the evolution of the intergalactic medium, and the abundance of metal absorbers. 
We discuss evidence that adjustments either to the assumed ratio of carbon and 
silicon yields or to the geometry of ionizing escape may be required.
We analyze the predicted relationship between galaxies and absorbers in
Section~\ref{sec:hosts}. Finally, we summarize in Section~\ref{sec:sum}.
 
\section{Simulation}\label{sec:sims}
Our simulation is an update to the \td~calculations described in~\citet{finl18}. 
It assumes the same cosmology in which 
$(\Omega_M, \Omega_\Lambda, \Omega_b, H_0, X_H) = (0.3089, 0.6911, 0.0486, 67.74, 0.751)$.
However, it incorporates several adjustments to dynamic range and subgrid 
physics that were motivated by discrepancies with observations as discussed 
there. Here, we outline those updates and discuss their effects.

\subsection{Adjustments to the Feedback Model}\label{ssec:calibrations}
Our newest calculation models a $15\hmpc$ volume with $2\times640^3$ 
mass resolution elements, and the UVB is modeled using $80^3$ spatial
resolution elements (``voxels"). This ``pl15n640RT80NF24" simulation 
treats roughly twice the cosmological volume as our previous best 
calculation~\citep{finl18} with the same mass and spatial 
resolution, enabling us to account more completely for rare, bright 
galaxies without compromising on our ability to capture faint galaxies 
and model the post-reionization LAF.

We scale down the rate at which star-forming galaxies eject gas and metals
by 0.2 dex with respect to~\citet{finl18}. This rate is governed by the 
mass-loading factor $\eta$, which is the ratio of the rate at which galaxies 
eject gas to their star formation 
rate. Previously, we adopted this parameter's dependence on stellar mass 
$\eta(M_*)$ from the high-resolution simulations of~\citet{mura15}
without adjustment. However,~\citet{mura15} note that the normalization of their 
published calibration carries an uncertainty of 0.2 dex, which roughly matches 
the discrepancy between our predictions and observations of the galaxy stellar 
mass and rest-frame UV LFs at $z\sim6$ (cf.\ Figures 3 and 4 
of~\citealt{finl18}). For our updated simulations, we therefore adopt
\begin{equation}\label{eqn:mlf}
\eta(M_*) = 0.63 \times 3.6\left(\frac{\mstar}{10^{10}\msun}\right)^{-0.35}.
\end{equation}
With this adjustment, the galaxies in our simulation produce slightly more 
stars and metals. They also produce more ionizing photons because we have not
changed the underlying emissivity model, which is based on a modified version
of {\sc Yggdrasil}~\citep{zack11} as described in~\citet{finl18}. In order that 
it predict roughly the same reionization history while also reproducing recent 
measurements of the mean transmission in the post-reionization 
LAF~\citep{bosm18}, we adjust our escape fraction model to
\begin{equation}\label{eqn:fesc}
\fesc(z) = 0.166 \left(\frac{1+z}{6}\right)^{2.65},
\end{equation}
and we cap $\fesc(z)$ at a maximum value $\fescmax=0.31$. 
This $\fesc(z)$ model is slightly lower than our previous high-resolution 
model at all redshifts, and it is consistent with recent observational
inferences. For example, it predicts $\fesc=0.057$ at $z=3$, which lies 
below the observationally inferred value for bright galaxies at that redshift 
($0.09\pm0.01$;~\citealt{stei18}). Likewise, it predicts $\fesc=0.13$ at 
$z=4.5$, consistent with the upper limit of 0.13 inferred from
observations of H$\alpha$ emission at $z=$4--5~\citep{bouw16}.
While these agreements support the emerging view that galaxies could readily 
have driven reionization and dominated the post-reionization 
UVB~\citep{robe10,robe15,bouw16,plan16b,fink19}, Equation~\ref{eqn:mlf} 
remains an assumption that must be tested through more detailed observations. 
As a starting 
point, $\fesc(z)$ is assumed to be energy-independent; we will explore 
relaxing this assumption below.

\section{Comparisons with Previous Results}\label{sec:newResults}

\subsection{Observations of Galaxies}
We now demonstrate that, with the adjustments described in 
Section~\ref{ssec:calibrations}, our new simulation yields improved 
agreement with observations of the galaxy stellar mass function, the 
history of reionization and the post-reionization LAF,
and the abundance of metals in the high-redshift CGM. In order to 
make these comparisons, we apply the methods for identifying simulated 
galaxies and modeling absorption in the IGM/CGM previously described 
in~\citet{finl18}, to which the reader is referred for details on 
post-processing. As in that work, galaxy luminosities $\MUV$ refer to 
the rest-frame 1500\AA luminosity and are computed as AB magnitudes.

\begin{figure}
\centerline{
\setlength{\epsfxsize}{0.5\textwidth}
\centerline{\epsfbox{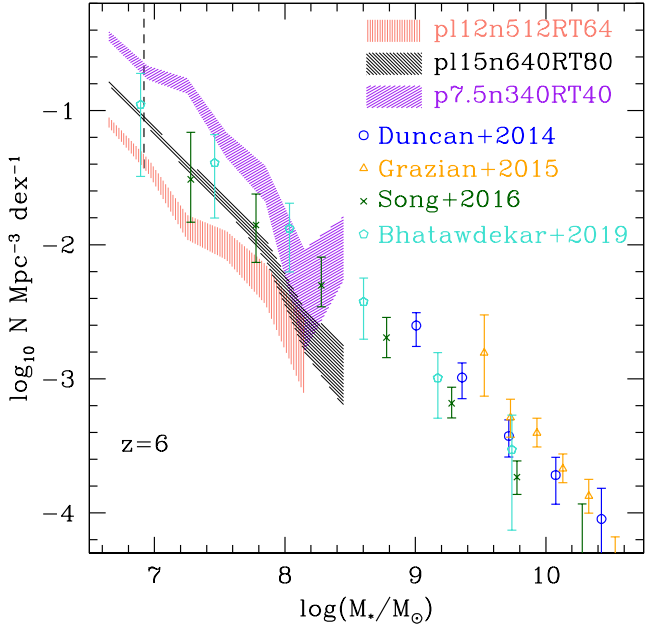}}
}
\caption{Stellar mass functions at $z=6$ in our previous and updated 
simulations versus observations; the pl15n640RT80 simulation (black)
represents our most up-to-date calibration.  The vertical tickmark 
indicates the 64-star particle mass resolution limit for all simulations. 
Turquoise pentagons are from~\citet{bhat19} and reflect their point-source 
incompleteness corrections; other points represent previous observations 
as indicated in the legend~\citep{dunc14,graz15,song16}.
}
\label{fig:mfz6}
\end{figure}

We begin with the galaxy stellar mass function (SMF) at $z=6$. In 
Figure~\ref{fig:mfz6}, we compare predictions from two 
previously-published simulations and our newest one versus observations. 
Comparing our previous calibration (pl12n512RT64NF24; magenta) versus our 
most recent one (pl15n640RT80NF24; black) reveals that suppressing 
outflows by 0.2 dex
(Equation~\ref{eqn:mlf}) boosts the predicted stellar mass of all galaxies 
by a similar factor, yielding improved agreement with the deepest available 
measurements. As our new simulation subtends nearly twice the cosmological 
volume, it also extends to slighly higher masses, slightly improving overlap 
with the observed dynamic range.

In an earlier work~\citep{finl16}, we presented a smaller, ``p7.5n340RT40"
simulation that readily reproduced the observed $\civ$ abundance. By comparing
with measurements of the $z\sim6$ SMF that have been carried out since that
work was published, we now see that, while it reproduced the observed $\civ$ 
abundance at $z>5$, it may have done so in part by overproducing stars and 
therefore metals~\citep{rahm16}.

The tendency for high-redshift star-forming
galaxies to be strong line emitters opens up the possibility of quantifying the
environments of high-redshift metal absorbers using narrow-line selection in
addition to broadband selections~\citep{cai17b}. As our simulations do not capture the 
detailed physics associated with emission and diffusion of Ly$\alpha$, we
model the outcome of these processes in post-processing via an empirical 
calibration.

\begin{figure}
\centerline{
\setlength{\epsfxsize}{0.5\textwidth}
\centerline{\epsfbox{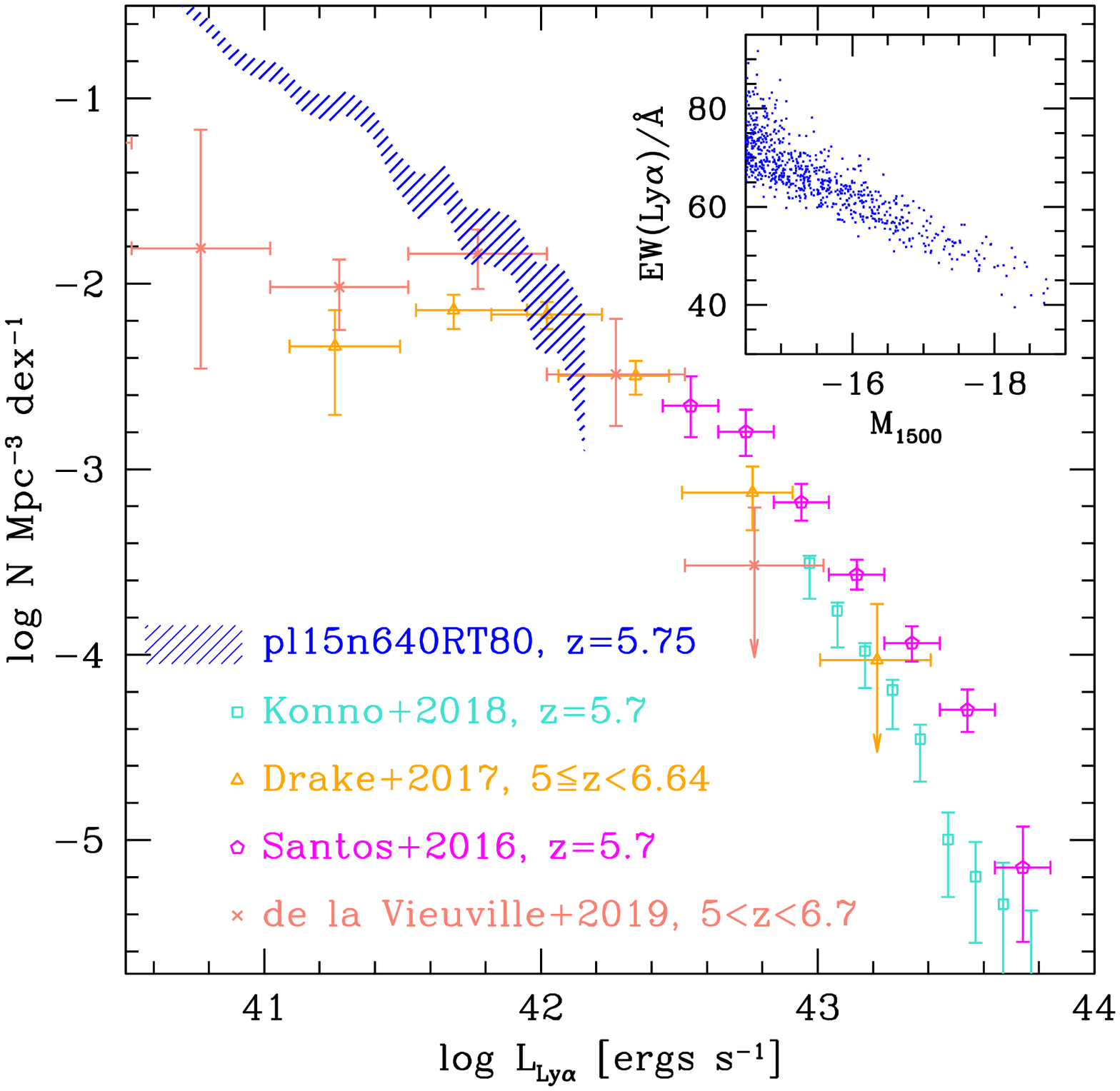}}
}
\caption{The shaded blue region shows the predicted Lyman-$\alpha$ luminosity 
function at $z=5.75$ (see text); its width indicates $\sqrt{N}$ uncertainty.
Orange triangles, turquoise squares, magenta pentagons, and salmon crosses
 indicate observations by~\citet{drak17},~\citet{konn18},~\citet{sant16},
and~\citet{vieu19}, respectively. All observations have been adjusted to our 
assumed cosmology. The predicted and observed LFs are in good 
agreement where observational incompleteness is not severe.
}
\label{fig:lyalaf}
\end{figure}

For each simulated galaxy, we obtain the expectation value of its Ly$\alpha$
equivalent width (EW$_{{\rm Ly}\alpha}$) from its stellar mass using 
Equation 22 of~\cite{oyar17}.\footnote{This step is largely an extrapolation 
because only a few of our simulated galaxies at $z=5.75$ have 
$\mstar>10^8\msun$, the range that dominates the~\citet{oyar17} observations.}
We show the resulting $\MUV$-EW$_{{\rm Ly}\alpha}$ relationship
in the inset panel of Fig.~\ref{fig:lyalaf}. The intrinsic scatter in 
$\MUV(M_*)$ generates scatter in EW$_{{\rm Ly}\alpha}(\MUV)$, 
but broadly the adopted values are in the range 40--80\AA. Observations are 
incomplete in the luminosity range spanned by the model. For example, 
Fig.\ 6 of~\citealt{drak17} indicates that their 
$z\approx(5,5.5,6.64)$ \lya LFs are 50\% complete at 
$\log(L_{{\rm Ly}\alpha}/\erg\ {\rm s}^{-1})=(42.0,42.3,42.7)$. Nonetheless, the satisfactory
agreement between the predicted and observed Lyman-$\alpha$ luminosity 
functions in Fig.~\ref{fig:lyalaf} is encouraging. As no \emph{ad hoc} 
calibration has been applied to this analysis, the tentative agreement
can be viewed as a test of the simulated UV LF and $\MUV-M_*$
relations, which, when combined with the observed dependence of the 
Ly$\alpha$ equivalent width on stellar mass at $3 < z < 4.6$, are what 
yield the prediction in Fig.~\ref{fig:lyalaf}.

\subsection{Observations of $\hi$ Reionization}
\begin{figure}
\centerline{
\setlength{\epsfxsize}{0.5\textwidth}
\centerline{\epsfbox{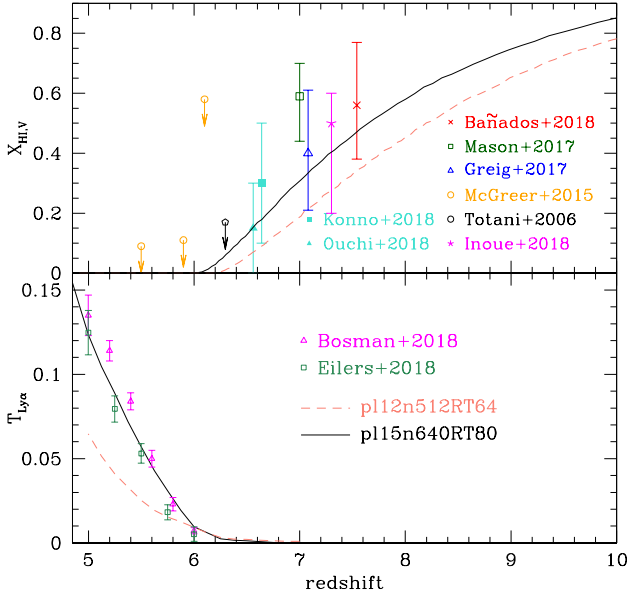}}
}
\caption{The history of reionization (top figure) and the mean transmission
in the Lyman-$\alpha$ forest in our previous and new simulations. Both are
in good agreement with the bulk of the measurements of the volume-averaged
neutral hydrogen fraction prior to overlap (top panel), but our most recent
calculation also yields a more realistic UVB in the post-overlap phase
(bottom panel).
}
\label{fig:HReion}
\end{figure}

The top panel of Figure~\ref{fig:HReion} shows that the updated simulation 
(pl15n640RT80NF24; 
solid black) predicts roughly the same overall reionization history as the 
previous one (pl12n512RT64; dashed salmon). In detail, reionization occurs
slightly later in the newer run, yielding improved agreement with
constraints on the pre-overlap neutral fraction, but the difference is not
large compared to uncertainties. By contrast, the bottom panel shows that
the predicted mean transmission in the LAF $\tla$ during the interval 
$5 < z < 6$ is in much better agreement with observations~\citep{bosm18,eile18}. 
The \emph{dis}agreement between our previous simulation
and observations of $\tla$ is unlikely to reflect primarily resolution limitations
because detailed convergence studies indicate that, for our choice of mass 
resolution and simulation volume, $\tla$ should be converged to 
$\leq 10\%$~\citep{dalo18,onor17}. We conclude that our newest simulation yields 
a UVB whose pre-overlap growth rate and post-overlap amplitude are suitably 
realistic for studies of the EOR CGM.

\subsection{Observations of Metal Absorbers}
\begin{figure}
\centerline{
\setlength{\epsfxsize}{0.5\textwidth}
\centerline{\epsfbox{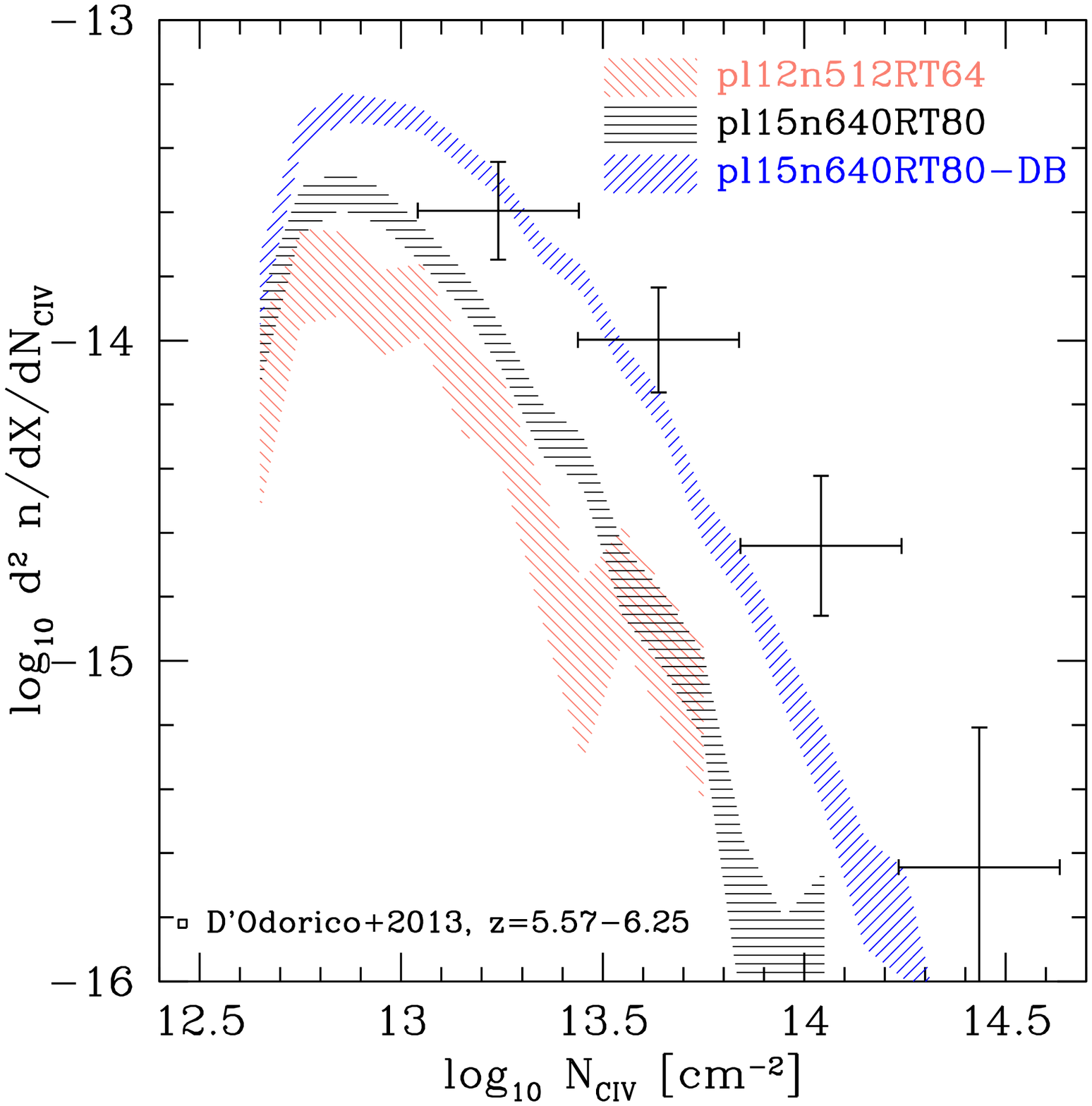}}
}
\caption{The predicted $\civ$ column density distribution in three different
scenarios versus observations.  The black, pl12n512RT64 region is an ``out-of-the-box"
representation of predictions from~\citet{finl18}. The salmon, pl15n640RT80 region
is an ``out-of-the-box" CDD compiled from our updated simulations. The blue,
pl15n640RT80-DB region shows that a density-bounded escape fraction model 
significantly boosts $\civ$. Observations are from~\citet{dodo13}.
}
\label{fig:dndXCIV}
\end{figure}

\begin{figure}
\centerline{
\setlength{\epsfxsize}{0.5\textwidth}
\centerline{\epsfbox{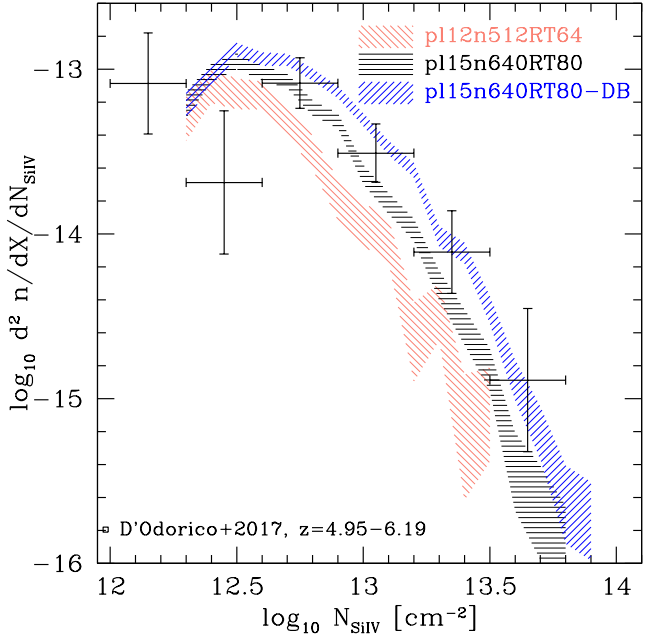}}
}
\caption{The predicted $\siiv$ column density distribution in three different
scenarios versus observations.  The black, pl12n512RT64 region is an ``out-of-the-box"
representation of predictions from~\citet{finl18}. The salmon, pl15n640RT80 region
is an ``out-of-the-box" CDD compiled from our updated simulations. The blue,
pl15n640RT80-DB region shows that a density-bounded escape fraction model 
boosts $\siiv$ production somewhat. Observations are from V.\ D'Odorico 
(private communication).
}
\label{fig:dndXSiIV}
\end{figure}

The improvements in Figures~\ref{fig:mfz6}--\ref{fig:HReion} allow us ask to what
extent small, observationally-permitted adjustments to the overall star formation 
efficiency and ionizing escape fraction could bring the $\siiv$ and $\civ$ CDDs 
predicted in our previous work into improved agreement with observations. 
We address these 
questions directly in Figures~\ref{fig:dndXCIV}--\ref{fig:dndXSiIV}.

Figure~\ref{fig:dndXCIV} confirms that, as expected, boosting the star formation
efficiency and the UVB amplitude does increase the overall $\civ$ production.
Note that this comparison benefits from a realistic 
treatment for observational incompleteness. We assume, following results from 
simulations by~\citet{dodo13}, that the observed $\civ$ census is (60, 70, 85, 100)\% 
complete at $\log(N_\civ/\cm^{-2}) = 13.3, 13.4, 13.5, 13.6$ and multiply the 
predicted abundance by this completeness function, interpolating to each column 
density.  Extrapolating this trend so that completeness falls to zero 
at $\log(N_\civ/\cm^2) \leq 12.6$ causes the predicted turnover at low 
columns.

While our new simulation does produce overall more $\civ$ than its 
predecessor, the predicted CDD still falls noticeably short of observations. 
Solving this problem by further boosting the UVB amplitude would compromise 
the excellent agreement with $\tla$ (Figure~\ref{fig:HReion}). Nor can the 
overall star formation efficiency be boosted, as this would overproduce the 
galaxy stellar mass function (Fig.~\ref{fig:mfz6}).

Could the offset indicate problems with the assumed metal yields?
In order to test whether an arbitrary adjustment to the overall metal 
yield\footnote{The overall metal yield is the ratio between the mass of new 
metals of all species ejected to the mass of long-lived stars 
formed~\citep{tins80}.} is indicated,
we compare in Figure~\ref{fig:dndXSiIV} the observed and simulated 
$\siiv$ CDDs at a similar redshift.  This comparison also accounts for
observational incompleteness: we scale the column
densities in the $\civ$ incompleteness function using the matched optical 
depth method so that the $\siiv$ completeness at any column density is
the same as the $\civ$ completeness at a column density that is a factor
2.432 higher. Taking this into account, we find that, whereas our previous 
simulation systematically underproduced $\siiv$, the updated 
prediction lies within $1\sigma$ of observations at all columns.

\begin{figure}
\centerline{
\setlength{\epsfxsize}{0.5\textwidth}
\centerline{\epsfbox{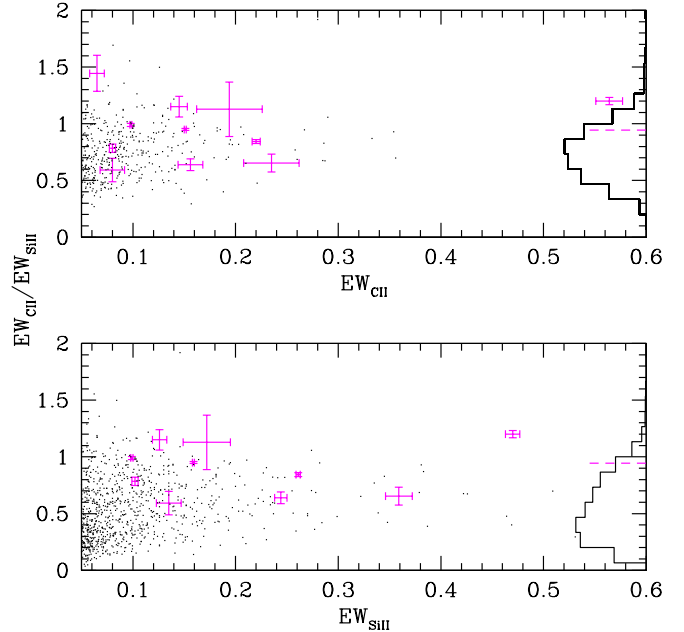}}
}
\caption{The distribution of $\cii/\siii$ equivalent widths versus $\cii$ 
(top) and $\siii$ (bottom) equivalent width in the simulation at $z=6$ (dots) 
versus in observations spanning $5.75 < z < 6.25$~\citep{beck19}. Histograms
collapse the predicted distribution along the x-axis after applying a 
minimum equivalent width of 0.05 \AA. The dashed segment indicates the 
weighted mean observed ratio. Both panels indicate that the simulated 
$\cii/\siii$ ratios are biased low.
}
\label{fig:EWrats}
\end{figure}

Figures~\ref{fig:dndXCIV}--\ref{fig:dndXSiIV} suggest that, if the 
assumed overall metal yield is correct, then our simulations cannot 
simultaneously match observations of the $\civ$ and $\siiv$ CDDs: 
boosting the UVB amplitude or the stellar mass density in order to 
match the $\civ$ CDD would lead us to overproduce $\siiv$. This may,
for example, contribute to the result discussed by~\citet{codo18}, 
whose reference simulation reproduced the observed high-redshift $\civ$ 
CDD but overproduced $\siiv$. We conclude that matching the observed 
$\siiv$ and $\civ$ CDDs simultaneously requires adjustments either 
to the \emph{relative} carbon and silicon yields, or to the slope 
of the UVB.

In order to explore the former possibility, we select co-patial $\cii$ and 
$\siii$ absorbers at $z=6$ in simulations and between $5.75 < z < 6.25$ in
the~\citet{beck19} observations and 
compare the distribution of $\cii/\siii$ equivalent width (EW) ratios in 
Figure~\ref{fig:EWrats}. As low-ionization absorbers tend to have weak 
ionization corrections~\citep{beck11a}, the EW ratio of co-spatial absorbers 
traces the underlying metal abundance ratio. The weighted 
mean observed EW ratio at $z\sim6$ is 0.94 (dashed segment in both panels). 
Selecting simulated systems using a cutoff of 0.05 \AA~in EW($\cii$,$\siii$), 
we find mean predicted ratios of (0.77, 0.54). Given that the measured 
abundance ratio of C/Si in the solar photosphere is 3.56~\citep{aspl09}, 
the model therefore agrees qualitatively with observations that the 
high-redshift C/Si abundance ratios are subsolar 
(see also~\citealt{beck11a}, Table 6). Quantitatively, however, they are
rather too subsolar; the predicted yield ratio of C to Si could be 
increased by 22-75\%. 

\begin{figure}
\centerline{
\setlength{\epsfxsize}{0.5\textwidth}
\centerline{\epsfbox{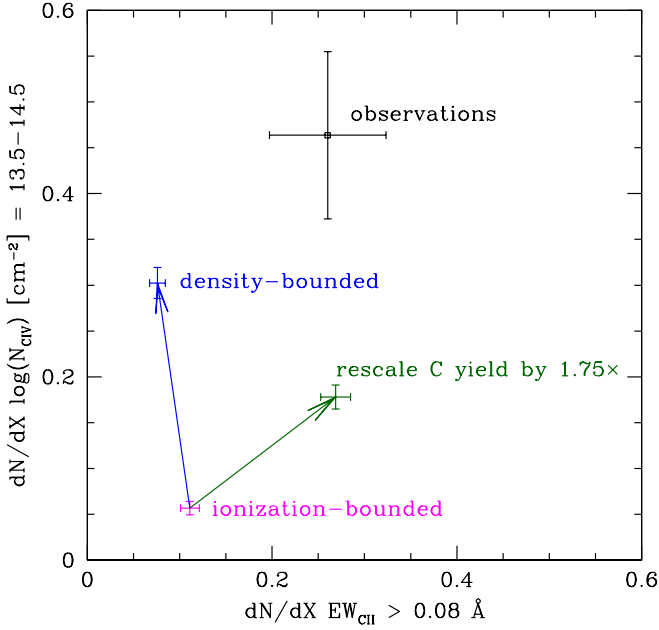}}
}
\caption{A comparison with joint constraints on the line incidences of 
$\cii$ and $\civ$. Observations of $\cii$ and $\civ$ are 
from~\citet{beck19} and~\citet{dodo13}, respectively. Without adjustment
in post-processing, the simulation underproduces both ions. Rescaling
simulated C abundances globally boosts $\cii$ into agreement with 
observations. A density-bounded escape model improves agreement with 
$\civ$ while exacerbating the $\cii$ underproduction.
}
\label{fig:dNdXCIICIV}
\end{figure}

Boosting the assumed carbon yields would increase the abundances of both
$\cii$ and $\civ$. As the tendency for low- and high-ionization ions
to evolve differently at $z>5$~\citep{beck11a,coop19} may encode key
insight into early IGM enrichment and reionization, we compare in 
Fig.~\ref{fig:dNdXCIICIV} the predicted and observed line incidences 
${\rm d}N/{\rm d}X$ of $\cii$ and $\civ$.  
For $\cii$, we adopt the catalog of low-ionization absorbers spanning
$z=5.75$--6.25 by~\citet{beck19} and select systems with EW$_\cii > 0.08$ 
\AA. This catalog probes an absorption path length of $\Delta X=65.34$ 
in our cosmology. For $\civ$, we obtain the observed line incidence of 
systems with column densities $13.5 < \log(N_\civ) < 14.5$ from the fit
to Fig.\ 19 of~\citet{dodo13}. Adjusted to our cosmology, this yields
$dN/dX(\civ) = 0.464\pm0.091$ for $5.3 < z < 6.2$. The simulated $\civ$
and $\cii$ line incidences are compiled using the same cuts in equivalent
width and column density. Incompleteness is modeled in $\civ$ as before,
while incompleteness in $\cii$ is modeled using a fit to Fig.\ 2 
from~\citet{beck19}. Uncertainties are $\sqrt{N}$ except in the case
of the observed $\civ$ line incidence, where we assume it is dominated
by the uncertainty in the slope of the observed $\civ$ column density
distribution.

Without adjustments in post-processing, the magenta point labeled
``ionization-bounded" in Fig.\ref{fig:dNdXCIICIV} indicates clearly
that the simulation underproduces
both $\cii$ and $\civ$. This supports the view that
the assumed metal yield from supernovae and/or evolved stars is too
low. Scaling all simulated C mass fractions up by a factor of 
$1.75\times$ (green arrow) brings the $\cii$ line incidence into 
agreement with observations while alleviating roughly half (in 
logarithmic units) of the $\civ$ discrepancy.

\subsection{A Model for Density-Bounded Escape}\label{ssec:db}

In this section, we ask whether physically-motivated changes to the way in 
which ionizing light escapes from galaxies could steepen the predicted UVB
enough to yield simultaneous agreement with $\siiv$ and $\civ$ observations.
While it is conventional to model the UVB under the assumption that the 
escape of ionizing photons from galaxies is an energy-independent scalar 
$\fesc$~\citep{brom01,haar12,khai19}, this model is not required 
\emph{a priori}. Physically, the assumption of a scalar $\fesc$ corresponds 
to a scenario in which ionizing flux escapes through transparent ``holes" 
in the interstellar medium (ISM) which are in turn separated by 
opaque ``walls"~\citep{wise09}. Equivalently, 
it may be imagined that a small fraction of massive stars lie 
``outside" the ISM~\citep{conr12}, although it has been argued that this 
particular effect may not dominate ionizing escape~\citep{kimm14}.
Following~\citet{zack13}, we refer to this scenario as the 
``ionization-bounded" escape model (IB). Direct evidence that some LyC 
flux escapes through optically-thin channels is provided by
LyC-leaking galaxies for which Ly$\alpha$ emission is observed at the 
systemic velocity~\citep{rive17,izot18b,vanz19}. This signature is
predicted in such scenarios theoretically~\citep{behr14}.

In order to explore the possible consequences of this widespread assumption, 
we consider the opposite extreme.  In this ``density-bounded" (DB) scenario, each 
star is separated from the ISM's boundary by a thin layer of gas whose column
density is tuned so that the overall escape of $\hi$-ionizing flux matches 
the IB model, but the optical depth to more energetic photons 
is lower.\footnote{Illustrations of these two escape scenarios are found in 
Figure 1 of~\citet{zack13} and~\citet{dunc15}.} Evidence that ionizing flux 
may escape from high-redshift galaxies via density-bounded media comes  
from observations that high [OIII]/[OII] emission line flux ratios and 
weak [SII] emission are reliable predictors of LyC 
emission~\citep{alex15,izot16,izot18a,flet19,vanz19}. Additionally,~\citet{stei18}
report that a representative sample of LyC-leaking galaxies at $z\sim3$ do not 
show appreciable Lyman-$\alpha$ emission at the systemic velocity.
We impose the DB scenario in post-processing as an adjustment to the UVB that 
is predicted on-the-fly. First, we compute the volume-averaged $\hi$ ionization 
rate $\langle\Gamma_\hi\rangle$:
\begin{equation}
\langle\Gamma_\hi\rangle \equiv \left\langle \int \frac{4 \pi J_\nu}{h \nu} \sigma_\nu \mathrm{d}\nu\right\rangle
\end{equation}
where $\langle\rangle$ indicate a volume-average and the other quantities have
their usual meanings. Figure~\ref{fig:HReion} insures that this ionization 
rate, which turns out to be $\scnot{3.37}{-13}\s^{-1}$ at $z=5.75$, is 
realistic.

Next, we rescale the simulated galaxy UVB by the reciprocal of the assumed 
$\fesc(z=5.75)=0.227$ (ie., $J_\nu \rightarrow J_\nu/\fesc$) and compute 
the $\hi$ column density that would return the same $\langle\Gamma_\hi\rangle$.  
This is accomplished by solving the equations
\begin{eqnarray}
\langle\Gamma_\hi\rangle & \equiv & \left\langle \int \frac{4 \pi J_\nu}{\fesc(z=5.75) h \nu} \exp(-\tau_\nu) \sigma_\nu \mathrm{d}\nu\right\rangle \\
\tau_\nu & = & \sigma_{\nu,\hi} N_\hi + \sigma_{\nu,\hei} N_\hei
\end{eqnarray}
for $N_\hi$. The obscuring gas column is assumed to be completely neutral 
with the same helium mass fraction as the simulation. The resulting $\hi$ column 
density turns out to be $\scnot{4.91789}{17}\cm^{-2}$. This low column density 
would manifest observationally in the Lyman-$\alpha$ emission profile as a small 
velocity offset and narrow peak separation~\citep{verh15,kaki19,kimm19}.
We then re-scale our simulated UVB by the ratio 
$J_\nu \rightarrow J_\nu \exp(-\tau_\nu)/\fesc(\nu)$  at all frequencies and 
positions. This adjustment is approximate because it assumes that
$J_\nu$ varies linearly with $\exp(-\tau_\nu)$.  In reality, increasing 
the emissivity decreases the opacity by ionizing more gas, which in turn 
amplifies the overall boost to $J_\nu$~\citep{mcqu11}.  The detailed, 
superlinear dependence of $J_\nu$ on the emissivity model can only be 
established via numerical simulations. Our heuristic model for 
DB escape is therefore conservative in the sense that
small changes to $\fesc(z,\nu)$ at high energies will have a larger 
impact than we estimate on $J_\nu$.

\begin{figure}
\centerline{
\setlength{\epsfxsize}{0.5\textwidth}
\centerline{\epsfbox{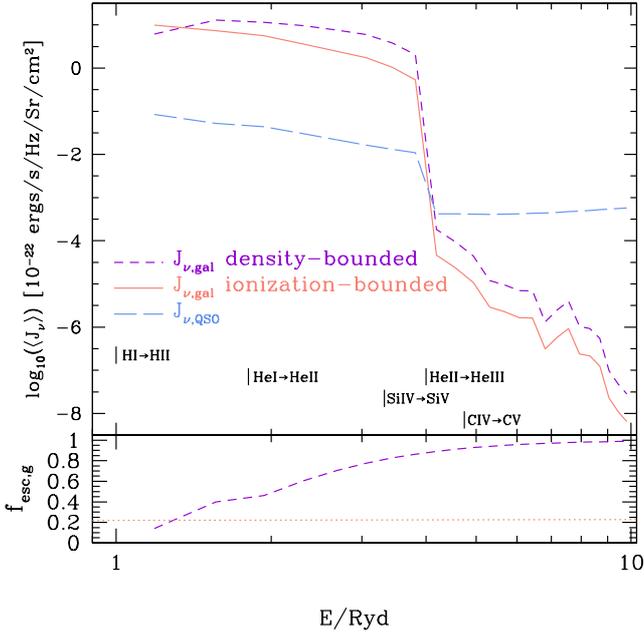}}
}
\caption{\emph{Top}: The volume-averaged galaxy and quasar UVBs. In 
an ``ionization-bounded" escape scenario (solid salmon), the galaxy UVB has 
an overall redder spectral slope than in a ``density-bounded" scenario 
(dashed purple) even though the two yield precisely the same overall $\hi$ 
ionization rate. \emph{Bottom}: In the ionization-bounded model, 
$\fesc=0.22681$. By contrast, $\fesc$ increases with energy in the
density-bounded scenario even though its emissivity-weighted mean 
is unchanged.
}
\label{fig:meanUVBs}
\end{figure}

We confirm in the top panel of Figure~\ref{fig:meanUVBs} that the volume-averaged 
galaxy UVB in the IB scenario has an overall steeper spectral slope than in the 
DB scenario. The difference is strongest for energies $<4$ Ryd, where $\fesc$
varies most strongly with energy in the DB scenario.
The top panel also indicates 
the ionization potentials for the ions that we consider. The impact on the 
LAF of switching between the two scenarios will be small by 
design, although a harder UVB will inevitably yield a qualitatively hotter IGM,
which may eventually be detectable in the LAF~\citep{boer19}. 
The impact on ions such as $\siiv$ and $\civ$, which are more sensitive to 
high-energy photons, will clearly be much more significant.

In the bottom panel of Figure~\ref{fig:meanUVBs}, we illustrate how the different 
spectral slopes in the top curve result from a different dependence of $\fesc$ on
energy. In the IB scenario, $\fesc$ is a constant, whereas in the DB
scenario it increases with energy, exceeding 50\% at 2 Ryd 
and rising above 90\% for $\heii$-ionizing energies ($ > 4 $ Ryd). In essence, the
DB model swaps low-energy flux for high-energy flux 
in order to achieve the same $\langle\Gamma_\hi\rangle$. 

We next ask whether the DB scenario can
reconcile simulations with observations of the statistics of $\civ$ and $\siiv$
absorbers. To do this, we re-extract simulated quasar sightlines that are identical 
to the ones considered previously except that, at all positions, the metal abundance
ratios are recomputed in the DB scenario. The quasar contribution 
to the UVB is retained without modification. We then identify and characterize 
synthetic metal absorbers in the same way as for the IB scenario.

The blue shaded region in Figure~\ref{fig:dndXCIV} shows that, by
increasing the $\civ$ fraction, the DB model roughly doubles each $\civ$ 
absorber's column density, nearly eliminating the $\civ$
discrepancy. Similarly, we show in Fig.~\ref{fig:dndXCIV} that
the DB model boosts the overall $\civ$ line incidence into near-agreement with
observations. This improvement comes at the expense of the $\cii$, whose line 
incidence is pushed further away from the observed range.\footnote{As the 
impact on $\siii$ is similar, the predicted distribution of $\cii/\siii$ 
equivalent with ratios (Figure~\ref{fig:EWrats}) is only weakly sensitive 
to the choice of escape fraction model.} The DB model's impact on $\siiv$ is 
weaker (Fig.~\ref{fig:dndXSiIV}) because $\siiv$ probes softer photons 
than $\civ$.

The comparisons in Figures~\ref{fig:dndXCIV}--\ref{fig:dndXSiIV} reiterate
that cosmological simulations confront grave difficulty in attempting to match 
simultaneously observations of the UVB, the galaxy stellar mass function,
and the $\civ$ and $\siiv$ CDDs at high redshift. Although the
$\civ/\siiv$ abundance ratio is a tracer of the UVB's spectral hardness that 
could constrain the relative contributions of galaxies and 
AGN~\citep{finl16,doug18}, these figures suggest that it could
alternatively probe the details of how ionizing light escapes from 
galaxies. In reality, as pointed out by~\citet{zack13}, the IB
and DB scenarios are opposite extremes and young stars will be 
separated from the CGM by sightlines spanning a distribution of column 
densities. Hence the question of how much $\civ$ is generated by light 
from galaxies versus quasars depends on the nature of this unknown column 
density distribution.

\begin{figure}
\centerline{
\setlength{\epsfxsize}{0.5\textwidth}
\centerline{\epsfbox{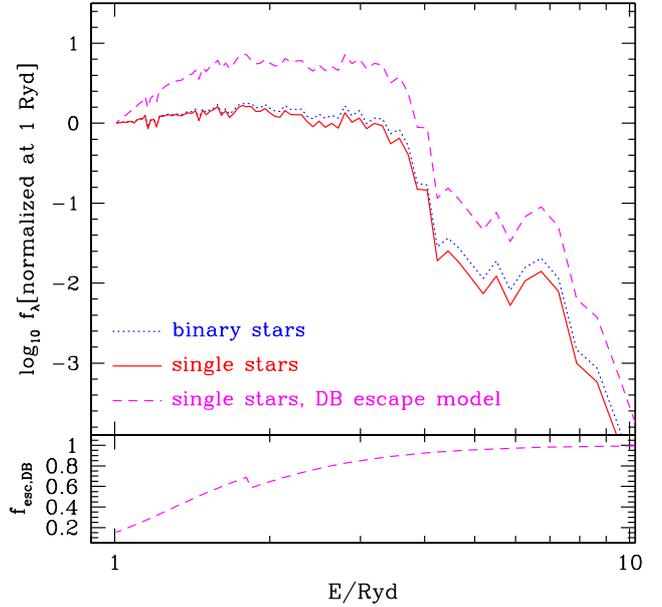}}
}
\caption{\emph{Top}: The emissivity from a stellar population that has been 
forming stars at a constant rate for 100 Myr.  The possibility of 
density-bounded escape is much more important at high energies than binary
stellar evolution. \emph{Bottom}: The escape fraction in the DB model.
}
\label{fig:bpass}
\end{figure}

This uncertainty is much more important for understanding high-ionization CGM
ions than the effects of binary stellar evolution, which have been shown to
increase the ionizing output of low-metallicity stellar populations and promote 
galaxy-driven reionization~\citep{stan16,ma16,rosd18}. The reason is that
spectral filtering associated with DB escape steepens the emerging ionizing 
continuum more than binary stellar evolution steepens the intrinsic one. To illustrate 
this point, we compare in Figure~\ref{fig:bpass} three hypothetical emerging 
spectra computed from version 2.2.1 of the Binary Population and Spectral 
Synthesis libraries~\citep{eldr17}. The red solid and blue dotted curves 
show spectra emerging from model galaxies that have been forming stars with 
$Z=0.002$ at a constant rate for $100\smyr$; the default initial mass 
function ``imf135\_300" is used in both cases. When normalized to 1 Ryd, the 
tendency for binary stars to produce more ionizing flux at high energies is 
reproduced. However, comparison with the dashed magenta curve shows that the 
effect is dwarfed by that of a purely DB escape
scenario: if we position the single-star model behind a neutral screen with a
hydrogen column density of $3\times10^{17}{\rm cm}^{-2}$ and account for
bound-free absorptions by both $\hi$ and $\hei$, then the escaping continuum 
steepens much more dramatically. In the bottom panel, we show the 
corresponding $\fesc(\nu)$, which rises from $\fesc=15\%$ at the Lyman limit 
to 100\% at 10 Ryd.

The uncertainty in the UVB joins other well-known uncertainties related to
the stellar population and to the possible role of dust. Decreasing the 
metallicity of stellar populations generically amplifies 
and hardens their Lyman continua~\citep{scha02}. Meanwhile, the predicted 
C/Si supernova yield ratio $y_{\rm C}/y_{\rm Si}$ in the~\citet{nomo06} 
models varies nonmonotically with metallicity, with an overall minimum of 
0.68 and a maximum of 1.59~\citep[][Table 3]{finl18}. The yield ratios are also 
sensitive at the $<15\%$ level to our assumption of a 50\% hypernova fraction 
(\emph{ibid.}). We have not explored the implications of varying the initial 
mass function (IMF), although it has been shown previously that a more 
top-heavy IMF can boost both C and Si yields~\citep{kulk13}. Finally, dust
extinction generically reddens the emerging UV continuum, potentially 
counteracting the spectral hardening in the DB scenario.

These considerations highlight two unexplored avenues for future research. 
First, as noted by~\citet{berg19}, it is worth revisiting high-resolution 
simulations of high-redshift galaxies in order to quantify how $\fesc$
varies with energy when both dust and a realistic distribution of ISM column 
densities are taken into account. Second, it is worth exploring how the 
predicted $J_\nu$ changes in synthesis models of the UVB~\citep{fauc09,haar12,khai19} 
if the conventional assumption of an IB escape scenario is relaxed.

\section{The Absorber-Galaxy Connection}\label{sec:hosts}
\subsection{Detecting Hosts in the Far-UV Continuum}

\begin{figure}
\centerline{
\setlength{\epsfxsize}{0.5\textwidth}
\centerline{\epsfbox{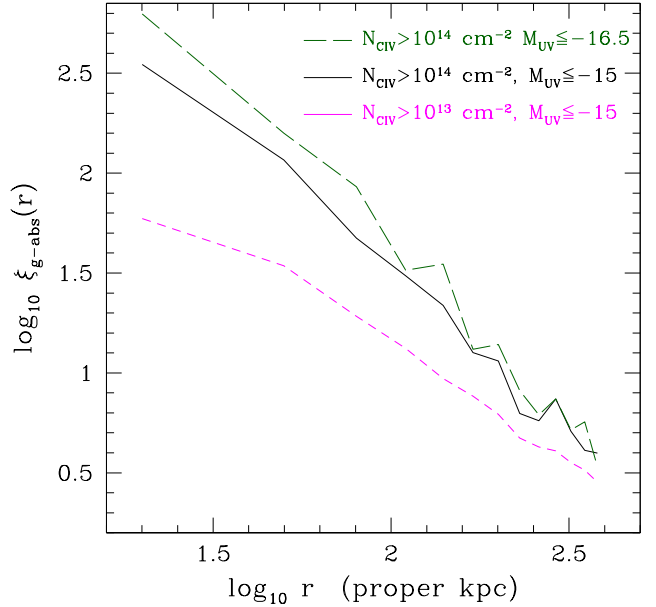}}
}
\caption{The $\civ$ absorber-galaxy cross-correlation function at $z=5.75$ in
the DB escape scenario. Line styles and colors correspond to different 
combinations of minimum column density and luminosity as indicated.
Galaxies and absorbers exhibit positive cross-correlation out to $>300$ 
physical kpc, and the amplitude of the correlation function increases 
independently with absorber strength and galaxy luminosity.
}
\label{fig:xi}
\end{figure}

Having established that our simulation produces the correct number of 
galaxies and a UVB whose amplitude at the Lyman limit is realistic, we now 
turn to our motivating question of how bright the neighboring galaxies of 
high-redshift $\civ$ absorbers are. We will show that the relationship 
depends on absorber strength, galaxy luminosity, distance, and the dependence 
of $\fesc$ on energy.
To enable this discussion, we identify simulated galaxies and compute 
their 1500\AA~continuum luminosities ($\MUV$) as described 
in~\citet[][\S 3.1]{finl18} and then match them in position space with the
simulated absorbers.
Next, we compute the galaxy-absorber cross-correlation 
function $\xiga(r)$ for different combinations of minimum absorber strength
and $\MUV$. The cross-correlation $\xiga(r)$ is computed as:
\begin{equation}\label{eqn:xi}
\xiga(r) \equiv \frac{1}{n_0}\frac{\Delta N(r)}{\Delta V} -1,
\end{equation}
where $\Delta N(r)$ is the mean number of galaxies within a finite spherical 
shell of volume $\Delta V$ located a distance $r$ from an absorber, and 
$n_0$ is the mean number density of galaxies averaged over the entire 
simulation volume. 
When computing $\xiga(r)$, we extract absorbers in the DB scenario because 
it reproduces the observed $\civ$ CDD best; alternative adjustments to
the model that boost $\civ$ would yield very similar results. We will use 
``abundance" and ``environment" interchangeably to refer to the local 
volume density of galaxies.

Figure~\ref{fig:xi} verifies that absorbers are positively correlated 
with galaxies out to distances of at least 300 proper kpc (pkpc). Moreover, the
amplitude of $\xiga(r)$ increases with both luminosity and 
absorber strength. In other words, stronger absorbers have more galaxies 
in their neighborhoods, and bright galaxies cluster more strongly 
about absorbers than faint ones do. 

This first hint that, even at early times, $\civ$ absorbers form more 
efficiently around brighter galaxies seems qualitatively inconsistent 
with observational results that the host galaxies of strong, high-redshift 
absorbers tend to be faint~\citep{diaz11,diaz14,diaz15,cai17b}. However, these 
results are not necessarily in tension: if the dependence of clustering 
strength on luminosity is weak whereas the slope of the luminosity
function's faint-end is steep, one still expects to find more faint 
galaxies than bright ones about strong absorbers.

Note that, in order to map from the absorber's position in velocity space 
to configuration space, we neglect its proper motion. For a typical velocity 
width of $\sim50\kms$, this simplification could blur the predicted 
galaxy-absorber relation on scales of up to $\sim75$ pkpc. If this effect 
were severe, it would cause $\xiga(r)$ to flatten at small $r$. The absence
of such a feature in Fig.~\ref{fig:xi} indicates that this blurring is not a 
serious issue.

\begin{figure}
\centerline{
\setlength{\epsfxsize}{0.5\textwidth}
\centerline{\epsfbox{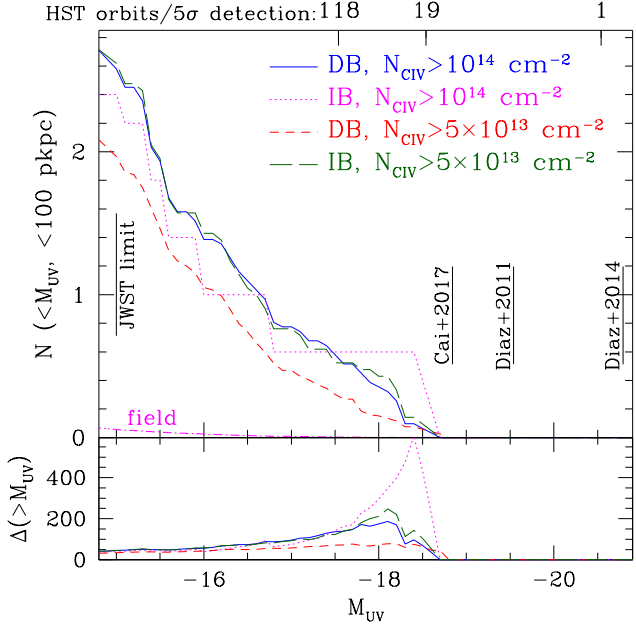}}
}
\caption{\emph{Top:} The predicted average number of detectable galaxies within 100 pkpc 
of a $\civ$ absorber at $z=5.75$ as a function of $\MUV$. Curves corresponding 
to different combinations of minimum $N_\civ$ and escape fraction scenario are 
identified in the legend. Irrespective of the column density or the choice of 
escape scenario, the
probability of detecting a single galaxy does not approach unity for surveys that
probe only to $\MUV\sim-17$, which is roughly \emph{HST}'s blank-field limit.
Vertical lines with references indicate the depths of published surveys. 
\emph{Bottom:}The local galaxy overdensity $\Delta\equiv N_\civ/N_{\rm field}$
as a function of luminosity.
}
\label{fig:hostCumCIV100}
\end{figure}

\begin{figure}
\centerline{
\setlength{\epsfxsize}{0.5\textwidth}
\centerline{\epsfbox{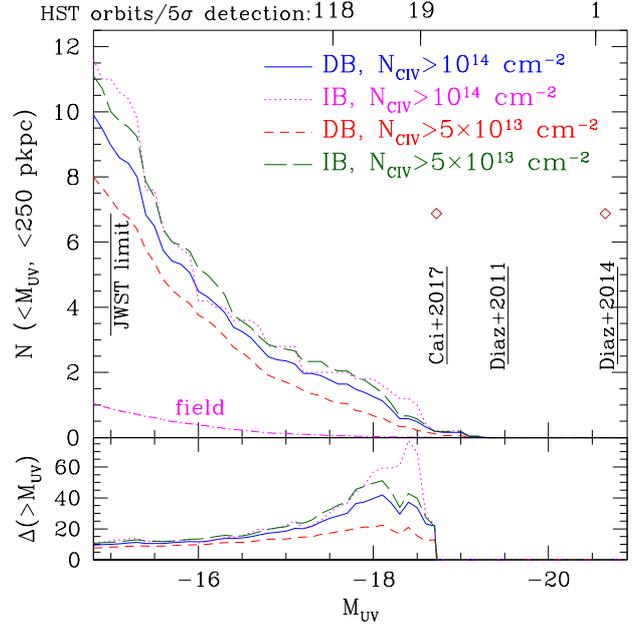}}
}
\caption{Same as Fig.~\ref{fig:hostCumCIV100}, but for a maximum radius
of 250 pkpc. More neighboring galaxies are predicted (top panel), but the 
mean galaxy overdensity out to this larger distance is lower (bottom panel).
Diamonds indicate the luminosities of candidate host galaxies that have been 
identified within 250 pkpc of a strong $\civ$ 
absorber at $z>4.5$~\citep{diaz14,diaz15,cai17b}; their position on the 
vertical axis is arbitrary.
}
\label{fig:hostCumCIV250}
\end{figure}

In order to explore how absorber environment may be quantified 
observationally, we group 
simulated galaxies that lie near absorbers (in three-dimensional space) into
two bins of minimum $\civ$ column density. Within each bin, we compute the 
mean cumulative number of galaxies that are located within 100 pkpc per 
absorber as a function of their minimum luminosity. For any given minimum 
luminosity, this number is an integral over the 
cross-correlation function $\xiga(r)$:
\begin{equation}
N(\leq \MUV, < 100\kpc) = \int_0^{100\kpc} 4\pi r^2 n_0 (1 + \xiga(r)) \mathrm{d}r
\end{equation}

\noindent
The solid blue curve in Fig.~\ref{fig:hostCumCIV100} shows the cumulative luminosity 
function of neighbors for absorbers with $N_\civ > 10^{14}\ \cm^{-2}$. As in 
Fig.~\ref{fig:xi}, the local galaxy abundance increases with absorber strength. 
Importantly, comparing the red short-dashed and green long-dashed curves
reveals that local galaxy abundance also depends on the escape fraction model:
if the escape fraction in the relevant energy regime is lower (as in the 
IB model), then more galaxies are expected per
$\civ$ absorber. This means measurements of absorbers' environments 
constrain the high-energy emissivity of faint galaxies.

In order to quantify how absorbers' environments vary with luminosity,
we use the dot-dashed magenta curve in the top panel to show the ``field" 
LF. This clearly indicates that galaxy abundance
is much lower in a randomly-selected region than in the vicinity 
of strong $\civ$ absorbers. By taking the ratio $\Delta$ of the absorber 
and field LFs, we recover the result from 
Fig.~\ref{fig:xi} that galaxy overdensity increases with luminosity. In 
particular, we show in the bottom panel of Fig.~\ref{fig:hostCumCIV100} 
that, within 100 pkpc, $\Delta$ rises from $\approx50$ for 
$\MUV=-15$ to $\approx 200$ for $\MUV = -18$. Future simulations 
incorporating a larger cosmological volume will be necessary to determine 
whether the predicted local overdensity continues to rise to brighter 
luminosities.

Even though overdensity increases with luminosity, most of the 
absorbers' neighbors are difficult to detect. The top axis of 
Fig.~\ref{fig:hostCumCIV100} converts the 1500\AA~luminosity to the 
number of \emph{HST} orbits required to detect a point source at 5$\sigma$ 
significance at $z=5.75$. We computed this conversion using the online Exposure 
Time Calculator\footnote{\url{http://etc.stsci.edu/etc/input/acs/imaging/}} for 
the Advanced Camera for Surveys F805LP ($z_{850}$) filter assuming that the source 
radiates as a 25 Myr Simple Stellar Population with 0.4 times solar metallicity.
Down to the \emph{HST} blank-field limit ($\MUV\sim -17$), fewer than one 
associated galaxy per strong absorber is expected within a 100 pkpc radius 
despite the strong association between galaxies and absorbers. This radius 
corresponds to 17" at $z=5.75$ in our cosmology, hence it is comparable to
the field of view of the Keck Cosmic Web Imager (KCWI) and smaller than the
or \emph{HST}/ACS field of view when imaging through its ramp narrowband 
filters. These instruments will not generally detect neighboring galaxies 
with a single pointing.

By contrast, we show in Figure~\ref{fig:hostCumCIV250} that a wider search 
radius---in this case, 250 pkpc---readily turns up more neighbors. The 
expected neighbor count does not vary linearly with the search area 
because galaxies are less clustered about absorbers at larger distances. 
This result was previously seen in Figure~\ref{fig:xi}, and it can be 
recovered by comparing the bottom panels of 
Figures~\ref{fig:hostCumCIV100}--\ref{fig:hostCumCIV250}: the mean 
overdensity for galaxies with $\MUV=-18$ falls from $\approx200$ within 
100 pkpc to $\approx40$ within 250 pkpc (see also Fig.~\ref{fig:xi}). 
A single deep \emph{HST}/ACS pointing (100--150 orbits in 
$z_{850}$) encloses this entire area and is predicted to uncover 2 
galaxies. Redshift confirmation would require equally deep imaging in 
additional filters, so this approach 
remains impractical. Fortunately, \emph{JWST} will soon prove extremely 
efficient at characterizing absorber environments: assuming a 
blank-field detection limit of $\MUV=-15$, \emph{JWST}/NIRSPEC will, 
with a single deep pointing, identify $\sim10$ neighboring galaxies 
per strong high-redshift $\civ$ absorber.

\citet{meye19} recently used an abundance-matching technique to estimate
the typical $\MUV$ and $M_h$ of $\civ$ hosts at $5.3 < z < 6.2$.
Assuming a one-to-one relation between $\civ$ absorbers and dark matter 
halos and a metal enrichment radius of 100 pkpc, they find that systems
with $\log(N_\civ) > 13$ are associated with $>10^{10}\msun$ halos and
$\MUV < -16$ galaxies. We may compare this estimate directly with our 
model: In the DB model, the minimum luminosity such that at least one
galaxy is contained within 100 pkpc of such an absorber at $z=5.75$ is
$\MUV < -14.5$, 1.5 magnitudes fainter than the~\citet{meye19} 
estimate. Although the two models differ in many ways, we suspect that
the primary difference is the halo-absorber relationship: in our model,
$\civ$ clouds from individual galaxies overlap in such a way that, 
although absorbers and galaxies are physically associated out to scales
of $\approx300$ pkpc (Fig.~\ref{fig:xi}), the geometric cross-section per
$M_h > 10^{10}\msun$ halo is smaller.  Equivalently, our model predicts
an abundant population of even fainter neighboring galaxies.

\subsection{Comparison with Observations}

To date, the published literature contains only a few robust associated 
galaxy-$\civ$ absorber pairs at $z>5$.  Here we summarize their properties;
where necessary we have converted to our simulation's cosmology.
Target 1 in~\citet{diaz11} confirms a galaxy with $\MUV=-20.66\pm0.05$, 
$L_{\mathrm{Ly}\alpha} = 8.01\pm0.55\times10^{42}\erg\ {\rm s}^{-1}$
lying at a transverse separation of 311.4 pkpc from the weak $\civ$ 
absorber identified as System 10 in Table A3 of~\citet{dodo13}.
~\citet{diaz14} search around two strong absorbers and identify 
a candidate neighboring galaxy as a Ly$\alpha$ emitter (LAE) 103027+052419, 
which has a tranverse separation from the $\civ$ absorber of 212.3 
pkpc~\citep[see also][]{diaz15}. Its UV continuum luminosity is
$\MUV=-20.65\pm0.52$. 
Finally,~\citet{cai17b} identified three LAE candidates around four 
strong $\civ$ absorbers. Two of these were not detected in 
more recent measurements using an integral field unit (IFU) 
spectrograph (D\'{i}az et al.\ 2020, \emph{in press.}),
but the candidate near the absorber at $z_\civ=4.866$ has not yet been 
followed up using other instrumentation, so for the present we assume 
it to be genuine. Its transverse separation from the
absorber is 210 pkpc, and its continuum and \lya line luminosities are 
$\MUV=-18.72\pm0.19$, and 
$L_{\mathrm{Ly}\alpha} = 2.60\pm0.53\times10^{42}\erg\ {\rm s}^{-1}$. 
In total, these efforts have surveyed the environments near six strong 
$\civ$ absorbers and identified three confirmed or candidate neighboring 
galaxies. Two of these lie within a transverse impact parameter of
250 pkpc of a strong absorber; their luminosities are indicated using 
diamonds in Figures~\ref{fig:hostCumCIV250} and~\ref{fig:hostCumLya250}.

Are these two neighbors expected? We cannot compare these statistics 
directly to our predictions owing to the fact that our simulation volume 
is not large enough to form galaxies brighter than $\MUV=-19$ at $z>5$. 
However, given that $\Delta$ varies slowly with luminosity, we may simply 
assume a fiducial $\Delta = 40$ (bottom panel of 
Figure~\ref{fig:hostCumCIV250}). Integrating the LF 
of~\citet[][Table 4]{fink15}, we find that the mean comoving space 
density at $z=5.75$ of galaxies brighter than $\MUV=(-19,-20)$ is 
$(9.25,2.93)\times10^{-4}\Mpc^{-3}$. We would therefore
expect an average of 0.74 $(\Delta/40)$ neighboring galaxies 
brighter than $\MUV = -19$ within 250 pkpc of each strong absorbers,
or 4.44 galaxies about the six absorbers for which positive detections
in follow-up surveys have been published to date. This order-of-magnitude 
level of agreement begs for improved constraints, which will be provided 
in the long term by observations using \emph{JWST} and in the short term 
by surveys with IFUs. 

\subsection{Detecting Hosts in Lyman-$\alpha$}
\begin{figure}
\centerline{
\setlength{\epsfxsize}{0.5\textwidth}
\centerline{\epsfbox{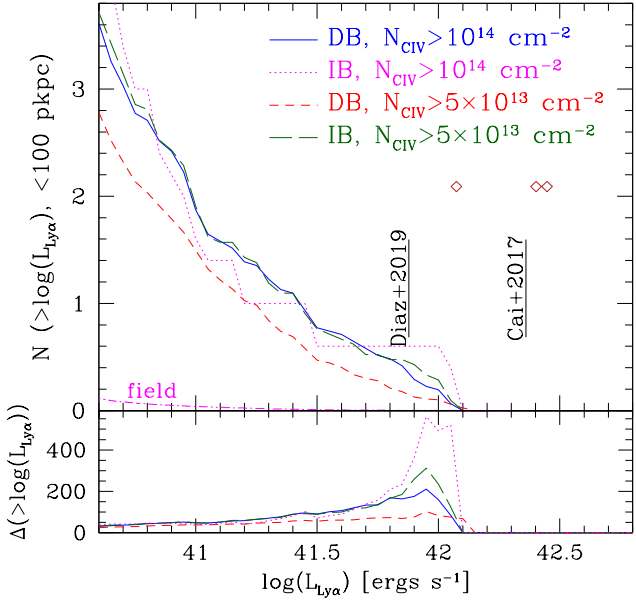}}
}
\caption{\emph{Top}: The predicted average number of detectable galaxies within 100 pkpc 
of a $\civ$ absorber at $z=5.75$ as a function of Ly$\alpha$ luminosity. Curves 
corresponding to different combinations of minimum $N_\civ$ and escape fraction 
scenario are identified in the legend. The vertical segment indicates the 
$4\sigma$ detection limit of~\citet{cai17b}, converted to our adopted 
cosmology. Diamonds indicate published observed neighboring 
galaxies (\citealt{diaz14,diaz15,cai17b} as well as systems matching
strong $\civ$ absorbers at $z>5.5$ that are identified in upcoming work~\citep{diaz20}.\emph{Bottom}: Local galaxy overdensity.
}
\label{fig:hostCumLya100}
\end{figure}

\begin{figure}
\centerline{
\setlength{\epsfxsize}{0.5\textwidth}
\centerline{\epsfbox{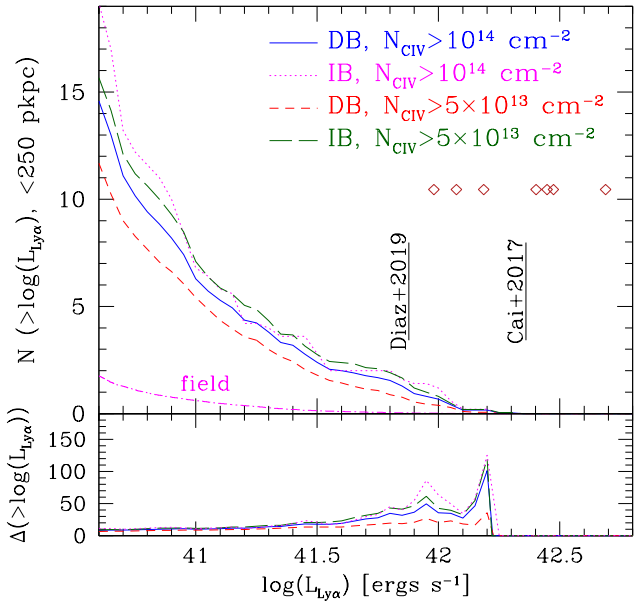}}
}
\caption{Same as Fig.~\ref{fig:hostCumLya100} but for a maximum radius 
of 250 pkpc.}
\label{fig:hostCumLya250}
\end{figure}

In Fig.~\ref{fig:hostCumLya100}, we show that, irrespective of the choice of UVB
or minimum $\civ$ column density, the neighboring galaxies of strong
$\civ$ absorbers are expected to be predominantly faint LAEs.
The predicted number of neighboring LAEs is sensitive to the absorber column
and UVB model in the same ways that emerge from Fig.~\ref{fig:hostCumCIV100}. 
Comparing the solid blue and short-dashed red curves reveals that the neighboring 
galaxies of stronger absorbers are brighter. Likewise, for a fixed column density, 
the density-bounded UVB predicts overall fainter galaxies because those absorbers 
are in this case associated with less rare systems. 

As in the case of stellar continuum detections, we find in the bottom panel of
Fig.~\ref{fig:hostCumLya100} that the ratio $\Delta$ of the predicted number density of 
LAEs that lie within 100 pkpc of an absorber to the field LF
is in the range of 100--500. In the case of the DB model and strong absorbers, 
$\Delta$ increases from $\approx50$ at the faintest luminosities currently probed 
($L_{\mathrm{Ly}\alpha}=10^{41} \erg\ {\rm s}^{-1}$) to 
$\approx200$ at the brightest luminosities captured by our simulation. Whether 
$\Delta$ continues to grow at higher luminosities can only be established
by simulations subtending larger cosmological volumes. It decreases with 
absorber strength and is overall higher in the IB model.
Finally, we confirm in Figure~\ref{fig:hostCumLya250} that a wider search area 
readily turns up more neighboring galaxies. In particular, a single deep 
VLT/MUSE pointing that probes to $10^{41.5}\erg\ {\rm s}^{-1}$ should 
uncover roughly three neighboring LAEs per strong $\civ$ absorber at $z=5.75$.

\subsection{Discussion: The Possible $\civ$-LAE Association}
Do strong $\civ$ absorbers prefer the company of faint galaxies or bright 
ones? While only a few candidate neighboring galaxies are currently 
identified in the stellar continuum within 300 pkpc of strong
$\civ$ absorbers at $z>5$, an intriguing result from~\citet{diaz14} was the suggestion
that, on scales of 10--20 comoving $\hmpc$, the surface density of (bright)
LBGs is systematically low in the vicinity of strong
$\civ$ absorbers while the surface density of LAEs
is systematically high. Given that LAEs are fainter in the stellar continuum
than LBGs, they speculated that $\civ$ absorbers may arise preferentially
in the vicinity of faint galaxies rather than bright ones. This is not 
expected. Figures~\ref{fig:xi}--\ref{fig:hostCumLya250} all indicate that 
the environments of strong $\civ$ absorbers are expected to be more overdense 
in brighter galaxies, the opposite to what~\citet{diaz14} infer.
Subject to the (significant) caveat that our simulation does not sample 
bright galaxies or large spatial scales adequately to address 
the~\citet{diaz14} observations directly, we may speculate qualitatively 
as to what the apparent disagreement suggests.

One interpretation is that the UVB's spatial fluctuations are in reality much
stronger than our model predicts. For example, we assume for simplicity 
that all galaxies have the same ionizing escape fraction $\fesc(z)$.
If, instead, $\fesc$ decreases with mass~\citep{alva12,paar15,bian17,stei18}, 
then the dependence of overdensity on luminosity could flatten, strengthening
the UVB amplitude and $\civ$ abundance in voids. However, it is not clear 
that the dependence could be reversed entirely given the tendency for faint 
galaxies to cluster about bright ones.

An alternative explanation invokes a coincidence whereby $\civ$ absorbers
and LAEs both arise preferentially in voids, though not for the same reason. 
Because $\civ$ is sensitive to $\heii$-ionizing photons, the portion of the 
UVB that regulates the $\ciii/\civ$ ratio remains highly inhomogeneous until 
$\heii$ reionization completes ($z<3$;~\citealt{mira98,mcqu16}). At earlier
times, it is likely that $\heii$ reionization is characterized by an 
outside-in topology such that overdensities are more opaque at high energies 
than mean-density regions or voids. This topology is often seen in numerical 
simulations of reionization~\citep{naka01,finl09b,katz17} and results from 
the tendency for ionization fronts to ``leak" out of overdensities into
voids. It is in fact predicted  by our simulations: 
at $z=5.75$, when the volume ionized fraction for $\heiii$ is 50\%, 
the ratio of the volume-averaged to the mass-averaged $\heiii$ 
fractions is 1.5, clearly indicating that voids are more transparent to
$\heii$-ionizing photons than overdensities. If this topology is correct, 
then the $\ciii/\civ$ ratio will inevitably be higher in overdense 
regions where LBGs are found. If, furthermore, $\ciii/\civ$ increases more
rapidly with overdensity than metallicity, then an anti-correlation 
between strong $\civ$ absorbers and LBGs is expected.

In order for this phenomenology to predict an association between $\civ$
absorbers and LAEs, it is necessary that the Ly$\alpha$ EW decreases with
increasing overdensity.~\citet{muld15} show that this could follow from 
quenching processes that occur predominantly in overdensities. Moreover, 
it is directly observed at $z\sim3$, where clustering measurements 
indicate that LBGs with Ly$\alpha$
in emission reside preferentially in the outskirts of overdensities and
in the field whereas LBGs that reside in overdensities have Lyman-$\alpha$
preferentially in absorption~\citep{cooke13}. Other studies have
found that narrowband-selected LAEs are less clustered than 
LBGs throughout $3 < z < 7$~\citep{ouch10,biel16}. Likewise,
studies of individual protoclusters at $3 < z < 6$ suggest that cluster 
members have
systematically weaker Ly$\alpha$ emission than field galaxies of
comparable luminosity (\citealt{tosh16,lema18}; see, however,~\citet{dey16},
who did not find this in the case of PC 217.96+32.3, which is at $z=3.786$).
Similarly,~\citet{cai17a} note that the large-scale distribution of LAEs 
in the BOSS1441 field shows a broad central core that 
is otherwise unexpected for a protocluster. 

A tendency for LAEs to avoid overdensities may also contribute to the 
result reported by~\citet{beck18}, who found an underdensity of 
LAEs centered about a sightline where the LAF is unusually optically 
thick at $z=5.7$~\citep{beck15a}. They attributed the LAE underdensity 
to the presence of a large-scale void based on modeling in which the 
\lya equivalent width was assumed to have no explicit environmental 
dependence~\citet{davi18c}. This conclusion received subsequent
support from an analogous survey that uncovered a large-scale 
deficit of LBGs at the same redshift along the~\citet{beck15a} 
sightline~\citep{kash19}. The emerging result that, whereas
regions where the LAF is unusually optically thick are associated with 
large-scale overdensities at $z=$2--3~\citep{cai17a}, they instead trace 
voids at $z>5$, may encode key insight into how galaxies drive the UVB's 
evolving spatial inhomogeneity. For the present discussion, however, 
the~\citet{beck15a} sightline may provide only ambiguous 
support for the hypothesis that LAEs selectively occupy voids.

In summary, if we assume that LAEs avoid overdensities at $z\sim6$, 
then they could
be physically associated with strong $\civ$ absorbers owing to a 
coincidence: During the outside-in stage of $\heii$ reionization, 
faint galaxies that cluster about LBGs are not visible as LAEs, and
their CGM is not visible in $\civ$. By contrast, voids host more of
the LAEs and enjoy higher $\heiii$ and lower $\ciii/\civ$ ratios 
owing to their spectrally harder UVB.

\section{Summary}\label{sec:sum}

We use the \td~simulations to study how systematic surveys 
for the neighboring galaxies of strong metal absorbers can constrain the 
details of the galaxy-driven reionization hypothesis. We reduce the mass flux 
in star formation-driven outflows by 37\% with respect
to~\citet{finl18} in order to improve agreement with observations of the galaxy
stellar mass function; this reduction is within the theoretical 
uncertainty quoted by~\citealt{mura15}. By adopting an empirical relationship between 
stellar mass and \lya equivalent width, we also recover the observed \lya 
LF at $z=5.75$, suggesting that the model predicts a
realistic $\mstar-\MUV$ relationship. Adjusting $\fesc(z)$ down then enables
excellent agreement with observations of the history of reionization and of 
$\tla(z)$. 

These calibrations yield satisfactory out-of-the-box agreement with the 
observed $\siiv$ CDD at $z=5.75$, but the observed $\civ$ CDD remains 
underproduced. Comparisons with joint observations of $\siii, \siiv, \cii$, 
and $\civ$ indicate that the lingering $\civ$ discrepancy in 
Fig.~\ref{fig:dndXCIV} cannot be eliminated by re-scaling the overall 
metal yields or stellar mass density. Nonetheless, scaling up the
carbon yield by a factor of $1.75\times$ simultaneously eliminates
disagreements with observations of $\cii/\siii$ equivalent width ratios and 
of the overall $\cii$ line incidence. It also alleviates but does not 
eliminate the $\civ$ deficit.

As an 
alternative, we replace in post-processing the traditional ionization-bounded 
escape scenario in which $\fesc$ is energy-independent with a density-bounded 
scenario in which $\fesc$ increases with energy. In a somewhat extreme case 
where $\fesc>80\%$ for energies above 4 Ryd, we find significantly more $\civ$ 
without overproducing $\siiv$ although in this case the $\cii$ deficit is 
exacerbated. 

Having established that our underlying model for galaxy growth and 
reionization is realistic even if $\civ$ remains enigmatic, we turn to the 
spatial association between galaxies and absorbers and find:
\begin{itemize}
\item Galaxies with $\MUV<-15$ are positively-correlated with $\civ$ 
absorbers out to distances of at least 300 pkpc.
\item The correlation strength increases independently with $\civ$
column density and with $\MUV$ and $L_{{\rm Ly}\alpha}$.
\item For both stellar continuum and \lya surveys, the mean expected number
of neighbors is larger for stronger absorbers, and at fixed absorber strength
it is larger under the assumption of the IB escape model because more 
galaxies are required to create a sufficiently hard local ionizing 
background.
\item The mean overdensity for galaxies within a few magnitudes of
$L_*$ falls from $\approx$200--300 within 100 pkpc of a strong absorber 
to $\approx$40--60 within 250 pkpc. 
\item The most abundant population of
neighboring galaxies is too faint to be accessible to \emph{HST},
but integral field units can already detect more than one galaxy 
per absorber in Ly$\alpha$. Detection will become routine with 
\emph{JWST}.
\end{itemize}

We propose that, if the physical association between LAEs and strong $\civ$ 
absorbers at $z>5$ is real, then it could owe to a coincidence rather than to 
the tendency for LAEs to dominate the UVB. In this scenario, $\heii$ reionization 
is more advanced in voids than in overdensities, with the result that the
CGM ionization state favors $\civ$ in voids and lower ionization states in
overdensities. The observed tendency for \lya equivalent widths to decrease 
with overdensity then leads naturally to a strong but coincidental 
association between LAEs and strong $\civ$ absorbers.  

Our results indicate that it would be interesting to survey 
the galaxy environments of high-redshift $\siiv$ absorbers for two reasons. 
First, their overall abundance is well-reproduced and relatively insensitive
to the choice of escape fraction model (Fig.~\ref{fig:dndXSiIV}). 
Second, $\siiv$ has a lower ionization potential than $\civ$. As such, it 
is a more direct probe than $\civ$ of the contribution that galaxies made 
to the latter stages of $\hi$ reionization.

Our work leaves a number of questions unanswered, among them:
\begin{itemize}
\item How does $\fesc$ vary with frequency? How does this dependence influence
the metagalactic UVB?
\item If calibrating our outflow model and $\fesc(z)$ to observations of 
the galaxy stellar mass function and the mean transmission in the 
LAF leads naturally to good agreement with the $\siiv$ CDD, does
the stubborn discrepancy with $\civ$ observations imply problems with the
metal yield model, binary stellar evolution effects, the relative 
contribution of AGN, or $\fesc$?
\item Does the dependence of galaxy overdensity on luminosity turn over or
continue to grow to the higher luminosities that our limited simulation
volume does not capture?
\item Do the topology of $\heii$ reionization at $z>5$ and the dependence 
of EW$_{{\rm Ly}\alpha}$ equivalenth width on overdensity predict the
association between LAEs and strong $\civ$ absorbers that was suggested 
by~\citet{diaz14}?
\end{itemize}

\section*{Acknowledgements}
Some simulations contributing to this work were run on NMSU's {\sc discovery} 
cluster; for technical advice and support we thank the NMSU ICT department. Other
calculations used the Extreme Science and Engineering Discovery Environment (XSEDE), 
which is supported by National Science Foundation grant number ACI-1548562; we
thank M.\ Tatineni for technical support with those efforts. We also thank V.\ Eijkhout
for his assistance with code optimization, which was made possible through the XSEDE 
Extended Collaborative Support Service (ECSS) program. KF
thanks G.\ Becker, T.\ Suarez Noguez, E.\ Ryan-Weber, C.\ Steidel,
X.-W.\ Chen, M.\ Prescott, and C.\ Doughty for helpful conversations and 
encouragement. We thank V.\ D'Odorico for sharing her $\siiv$ measurements. 
KF also thanks the University of Wisconsin Astronomy Department for hosting him 
during July 2019. We thank the anonymous referee for many helpful suggestions that 
improved the paper.  
Our work made use of the WebPlotDigitizer tool (https://automeris.io/WebPlotDigitizer),
for which we thank A.\ Rohatgi, as well as E.~L.\ Wright's online cosmology
calculator~\citep{wrig06}.  This research would have been quite
unthinkable without the NASA Astrophysics Data System and the arXiv eprint 
service. The Cosmic Dawn Center is funded by the Danish National Research 
Foundation.

\section*{ORCID IDs}
Kristian Finlator \url{https://orcid.org/0000-0002-0496-1656}\\
C.\ Gonzalo D{\'{\i}}az	\url{https://orcid.org/0000-0001-5146-1358}\\
Zheng Cai \url{https://orcid.org/0000-0001-8467-6478}\\



\bibliographystyle{mnras}
\bibliography{civ} 

\setcounter{subsection}{0}
\appendix
\renewcommand{\thefigure}{A.\arabic{figure}}
\section*{Appendix}
\renewcommand{\thesubsection}{\Alph{subsection}}
\subsection{Resolution Convergence}\label{app:resCon}
A number of numerical studies have explored the possibility that the CGM's
metals can transfer between enriched and unenriched regions via small-scale
diffusion processes~\citep{shen10,hafe19}. For example, diffusion can strip 
metals from winds before they escape the ISM, decreasing the mass of metals 
that enters the CGM. Subsequently, it can mix metals from winds into the 
ambient CGM, boosting its cooling rate. Our simulation does not directly 
resolve processes that occur on scales smaller than $600 h^{-1} \pc$ (comoving).
While we have not implemented a treatment for metal diffusion, we argue 
that doing so would not change on our predictions.

First, we limit our study to strong high-ionization absorbers. Previous work 
has shown that these systems trace gas whose density is between 
10--1000$\times$ the cosmological mean~\citep[][Fig.~8]{rahm16}. 
At these densities, the impact of metal diffusion is expected to be 
limited, with the typical metallicity changing by 0.1--0.2 
dex~\citep[][Fig.~8]{shen10}. This is smaller than many of the other 
uncertainties mentioned in Sec.~\ref{ssec:db} including the choice of metal 
yield and escape fraction model.

Second, we have used a simple numerical convergence test to quantify overall
resolution limitations. We have modeled a $15\hmpc$ volume
using three simulations that are identical except that they adopt gas
particle masses $M_g$ of $\scnot{3.2}{7}$, $\scnot{4.0}{6}$, and 
$\scnot{5.0}{5}\msun$. The corresponding gas softening lengths $h_{\rm SPH}$ 
are 2.34, 1.17, and 0.59 $\hkpc$ (comoving). For consistency, all simulations 
adopt the spatially-homogeneous~\citet{haar12} UVB. Decreasing $M_g$ 
generically boosts early star formation~\citep{spri03b}, increasing the 
cosmically-averaged metallicity at early times. Decreasing $h_{\rm SPH}$ 
improves the accuracy with which small-scale mixing processes are treated.
For our purposes, the most important question is whether the $\civ$ and 
$\siiv$ CDDs converge. 

\begin{figure}
\centerline{
\setlength{\epsfxsize}{0.5\textwidth}
\centerline{\epsfbox{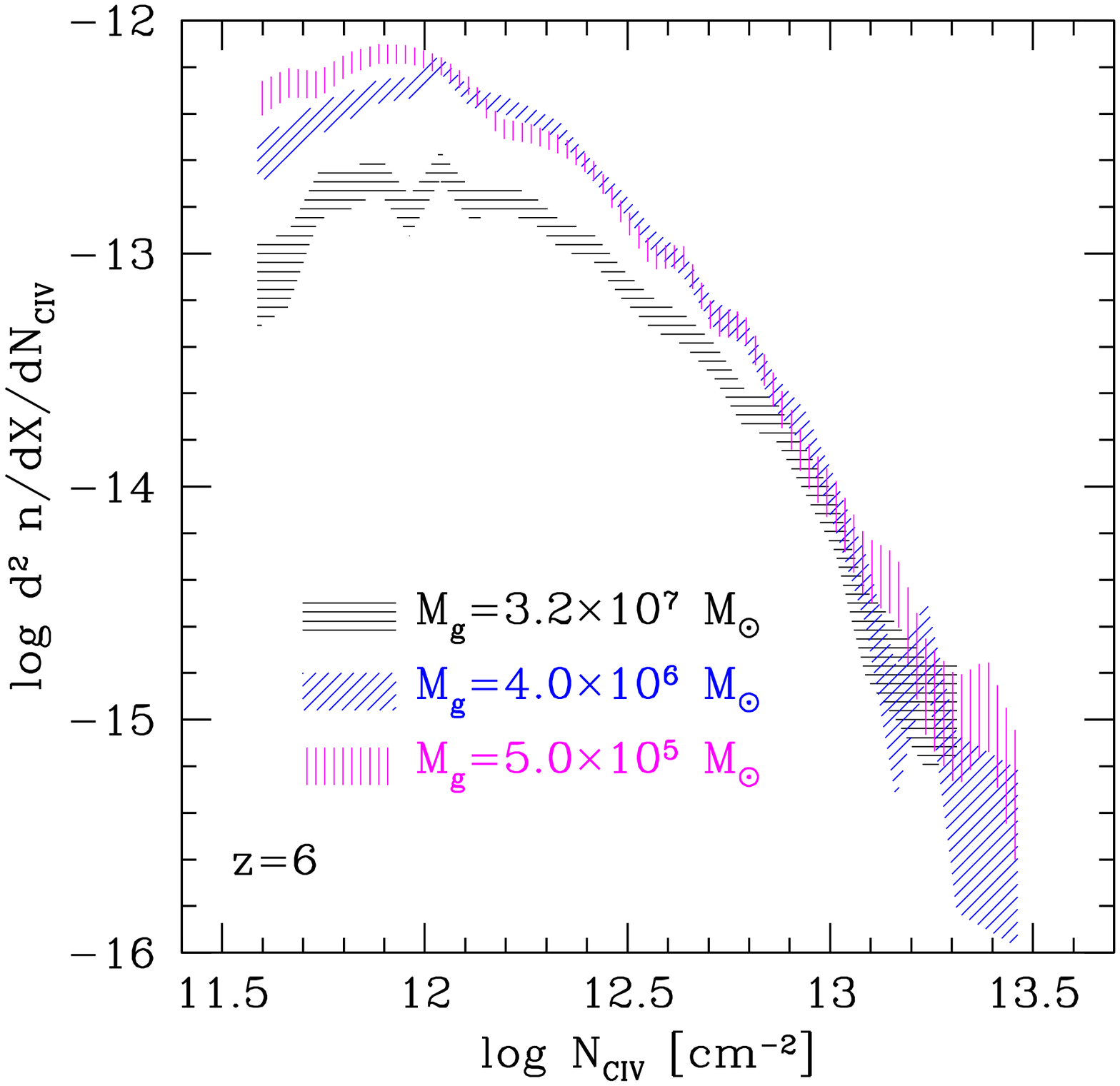}}
}
\caption{Resolution convergence test for $\civ$ CDD at $z=6$. 
For $N_\civ>10^{12}\cm^{-2}$, convergence requires a gas 
particle mass of $\scnot{5}{5}\msun$.}
\label{fig:resConCIV}
\end{figure}

\begin{figure}
\centerline{
\setlength{\epsfxsize}{0.5\textwidth}
\centerline{\epsfbox{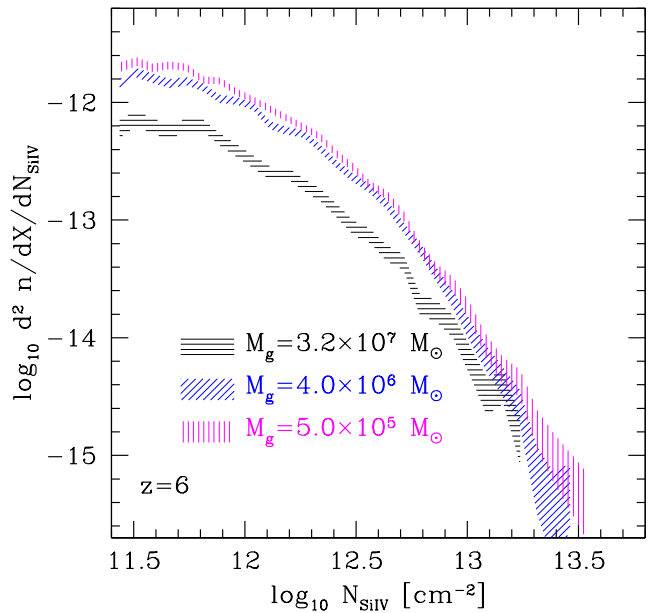}}
}
\caption{Resolution convergence test for $\siiv$ CDD at $z=6$. 
For a gas particle mass of $\scnot{5}{5}\msun$, the CDD
is converged to 0.1 dex for all $N_\siiv > 10^{12}\cm^{-1}.$}
\label{fig:resConSiIV}
\end{figure}

In Fig.~\ref{fig:resConCIV}, we show that the $\civ$ CDD converges
for $\civ$ column densities above $10^{12}\cm^{-2}$ as long as
$M_g < \scnot{5}{5}\msun$. At lower
columns, further enhancements to the mass resolution increase the 
predicted line incidence. Such relatively diffuse systems are 
also more sensitive to subgrid diffusion, but we do not consider
them in this work. In Fig.~\ref{fig:resConSiIV}, we show that the 
$\siiv$ CDD is likewise converged to within 0.1 dex at all columns
if the gas resolution satisfies
$M_g < \scnot{5}{5}\msun$. The simulation from which we draw this 
paper's predictions has a mass resolution of 
$M_g = \scnot{2.6}{5}\msun$, which is higher than the three test 
cases considered in Figs.~\ref{fig:resConCIV}--\ref{fig:resConSiIV}. 
For both of these reasons, we believe that our predictions are 
insensitive to small-scale metal diffusion.

\subsection{Cosmic Variance}\label{app:cosVar}
Our predictions may be subject to uncertainty owing to missing large-scale 
UVB spatial fluctuations. Fundamentally, they reflect the relationships 
between galaxies, their CGM, and the UVB. The physical association between 
galaxies and metals is determined by inflow and feedback processes that occur 
on length scales of up to $\sim$a virial radius. For a $10^{10}\msun$ dark matter 
halo at $z=6$, this is $\sim10$ pkpc, which is several hundred times smaller 
than our simulation volume. Spatial correlation between galaxies and metals
exists on scales of up to 300 pkpc owing to the underlying clustering of their
host galaxies, but even this scale spans $<10\%$ of our simulation 
volume for $z<6$.

The ionization state of those metals, by contrast, is dominated by the UVB, 
which fluctuates on scales up to roughly the mean free path.
At the Lyman limit, the mean free path $\lambda_{\rm mfp}$ is $10.0\pm1.5$ 
pMpc in our cosmology at $z=5.16$~\citep{wors14}, or roughly three times as 
large as our simulation volume.~\citet{wors14} find that $\lambda_{\rm mfp}$
evolves as $(1+z)^{-5.4\pm0.4}$ during $2.3<z<5.5$. Extrapolating suggests 
that the simulated UVB misses large-scale fluctuations for all $z < 7$ 
(see also~\citealt{ilie14}). The mean free path at higher energies is 
unconstrained. It is expected to increase with energy 
up to the $\heii$ edge and then drop sharply prior to the completion of
$\heii$ reioniation, with the detailed energy dependence reflecting the 
relative contributions of optically-thin gas versus Lyman limit systems.
As reionization proceeds, large-scale UVB fluctuations must weaken the 
relationship between a galaxy and the ionization state of its CGM because
the latter is increasingly influenced by light from distant sources rather
than from the local environment. Whether this erases the predictive power
of the galaxy-absorber relationship merits continued study, but even at 
$z=3$ it has been argued that the local environment dominates the 
ionization state of self-shielded systems~\citep{scha06}. We therefore
expect that galaxy environments about high-ionization metal absorbers 
will remain sensitive to $\fesc$ even within models that treat a larger
dynamic range.

\bsp	
\label{lastpage}
\end{document}

%% file: mymacros_nojournals
\newcommand{\hi}{\mathrm{H\,I}}
\newcommand{\hei}{\mathrm{He\,I}}
\newcommand{\heii}{\mathrm{He\,II}}
\newcommand{\heiii}{\mathrm{He\,III}}

\newcommand{\cii}{\mathrm{C\,II}}
\newcommand{\ciii}{\mathrm{C\,III}}
\newcommand{\civ}{\mathrm{C\,IV}}
\newcommand{\siii}{\mathrm{Si\,II}}
\newcommand{\siiv}{\mathrm{Si\,IV}}

\newcommand{\erg}{{\rm erg}}

\newcommand{\cm}{{\rm cm}}

\newcommand{\pc}{{\rm pc}}
\newcommand{\kpc}{{\rm kpc}}
\newcommand{\Mpc}{{\rm Mpc}}

\newcommand{\smyr} {\msun \mbox{ yr}^{-1}}

\newcommand{\lya}{Ly$\alpha$\ }
\newcommand{\hkpc}{h^{-1}{\rm kpc}}
\newcommand{\hmpc}{h^{-1}{\rm Mpc}}

\newcommand{\kms}{\;{\rm km}\,{\rm s}^{-1}}
\newcommand{\msun}{M_{\odot}}
\newcommand{\mstar}{M_*}
\newcommand{\MUV}{M_{\rm UV}}

\newcommand{\s}{{\rm s}}

\newcommand{\fesc}{f_{\rm esc}}

\newcommand{\scnot}[2]{#1\times10^{#2}}